%% file: cameraready.tex
\documentclass{LMCS}

\def\doi{8 (3:03) 2012}
\lmcsheading%
{\doi}
{1--31}
{}
{}
{Oct.~14, 2011}
{Aug.~10, 2012}
{}

\usepackage{enumerate}
\usepackage{hyperref}
\usepackage{amsmath,amsfonts}
\usepackage{amscd}
\usepackage{latexsym}
\usepackage{xspace}
\usepackage{color}
\usepackage{epsfig}
\usepackage{paralist}
\usepackage{rotate} 
\usepackage{graphicx}
\usepackage{booktabs}
\usepackage{proof}
\usepackage{wrapfig}
 
\newcommand{\defas}{\ensuremath{\stackrel{\text{\tiny def}}{=}}\xspace}
\newcommand{\iprincess}{\textsc{iPrincess}\xspace}
\newcommand{\opensmtitp}{\textsc{InterpolatingOpenSMT}\xspace}
\newcommand{\smtinterpol}{\textsc{SmtInterpol}\xspace}
\newcommand{\mathsat}{\textsc{MathSAT}\xspace}
\newcommand{\mathsatcongr}{\mathsat-\textsc{modEq}\xspace}
\newcommand{\mathsatceil}{\mathsat-\textsc{ceil}\xspace}
\newcommand{\dpll}{\textsc{DPLL}\xspace}
\newcommand{\sref}[1]{\S{}\ref{#1}}
\newcommand{\utvpi}{\ensuremath{\mathcal{UTVPI}}\xspace}
\newcommand{\euf}{\ensuremath{\mathcal{EUF}}\xspace}

\newcommand{\dl}{\ensuremath{\mathcal{DL}}\xspace}
\newcommand{\smt}{SMT\xspace}
\newcommand{\satres}{\textsc{sat}\xspace}
\newcommand{\unsatres}{\textsc{unsat}\xspace}
\newcommand{\larat}{\ensuremath{\mathcal{LA}(\mathbb{Q})}\xspace}
\newcommand{\laint}{\ensuremath{\mathcal{LA}(\mathbb{Z})}\xspace}
\newcommand{\smtt}{\ensuremath{\text{SMT}(\T)}\xspace}
\newcommand{\smttt}[1]{\ensuremath{\text{SMT}(#1)}\xspace}
\newcommand{\smtlaint}{\smttt{\laint}}
\newcommand{\T}{\ensuremath{\mathcal{T}}\xspace}
\newcommand{\Tmodels}{\models_{\T}}
\theoremstyle{plain}\newtheorem{property}[thm]{Property}
\newcommand{\efrac}[2]{\displaystyle\genfrac{}{}{0pt}{}{#1}{#2}}
\begin{document}

\title[Efficient Interpolant Generation in SAT Modulo \laint]{Efficient Interpolant Generation\\
  in Satisfiability Modulo Linear Integer Arithmetic}

\author[A.~Griggio]{Alberto Griggio\rsuper a}
\address{{\lsuper a}Fondazione Bruno Kessler, Trento, Italy}	%
\email{griggio@fbk.eu}
\thanks{{\lsuper a}Supported by Provincia Autonoma di Trento and 
the European Community's FP7/2007-2013 under grant
agreement Marie Curie FP7 - PCOFUND-GA-2008-226070 ``progetto Trentino'',
project ADAPTATION}

\author[T.~T.~H.~Le]{Thi Thieu Hoa Le\rsuper b}	%
\address{{\lsuper{b,c}}DISI, University of Trento, Italy}	%
\email{\{hoa.le, rseba\}@disi.unitn.it}  %

\author[R.~Sebastiani]{Roberto Sebastiani\rsuper c}	%
\thanks{{\lsuper c}Supported by SRC under
GRC Custom 
Research Project  2009-TJ-1880 WOLFLING
and under GRC Research Project 2012-TJ-2266 WOLF}	%

\keywords{Craig Interpolation, Decision Procedures, SMT}
\subjclass{F.4.1}

\begin{abstract}
\noindent The problem of computing Craig interpolants in SAT and SMT has
recently received a lot of interest, mainly for its applications in
formal verification.
Efficient algorithms for interpolant generation have been presented
for some theories of interest ---including that of equality and
uninterpreted functions (\euf), linear arithmetic over the rationals
(\larat), and their combination--- and they are successfully used
within model checking tools.
For the theory of linear arithmetic over the integers (\laint),
however, the problem of finding an interpolant is more challenging,
and the task of developing efficient interpolant generators for the
full theory \laint is still the objective of ongoing research.

\noindent In this article we try to close this gap.  We build on previous work and
present a novel interpolation algorithm for SMT(\laint), which
exploits the full power of current state-of-the-art SMT(\laint)
solvers. 
We demonstrate the potential of our approach 
with an extensive experimental evaluation 
of our implementation of the proposed algorithm in the \mathsat SMT solver.
\end{abstract}

\maketitle

\section{Introduction}
\label{sec:intro}

Given two formulas $A$ and $B$ such that $A\wedge B$ is inconsistent,
a {\em Craig interpolant} (simply ``interpolant'' hereafter) for
$(A,B)$ is a formula $I$ s.t.  $A$ entails $I$, $I\wedge B$ is
inconsistent, and all uninterpreted symbols of $I$ occur in both $A$
and $B$.

Interpolation in both SAT and SMT has been recognized to be a
substantial tool for formal verification.
For instance, in the context of software model checking based on
counter-example-guided-abstraction-refinement (CEGAR) interpolants of
quantifier-free formulas in suitable theories are computed for
automatically refining abstractions in order to rule out spurious
counterexamples.
Consequently, the problem of computing interpolants in SMT has
received a lot of interest in the last years 
(e.g., 
\cite{%
mcmillan_interpolating_prover,%
rybalchenko_interp,%
musuvathi_interpolation,%
interpolation_data_structures,%
tocl_interpolation,%
jain_cav08,%
lynch_interpolation,%
tinelli_tacas09,%
tinelli_cade09,%
ijcar10_interpolation,%
lpar10_interpolation%
}).
In the recent years, efficient algorithms and tools for interpolant
generation for quantifier-free formulas in SMT have been presented for
some theories of interest, including that of equality and
uninterpreted functions (\euf)
\cite{mcmillan_interpolating_prover,tinelli_tacas09}, linear arithmetic
over the rationals (\larat)
\cite{mcmillan_interpolating_prover,rybalchenko_interp,tocl_interpolation},
{fixed-width bit-vectors \cite{kroening_interp,griggio-fmcad11},}
and for  combined theories
\cite{musuvathi_interpolation,rybalchenko_interp,tocl_interpolation,tinelli_cade09},
and they are successfully used within model-checking tools. 

\smallskip
For the theory of linear arithmetic over the \emph{integers} (\laint),
however, the problem of finding an interpolant is more challenging.
In fact, 
it is not always possible to obtain quantifier-free interpolants
starting from quantifier-free input formulas
in the standard signature of \laint 
(consisting of Boolean connectives, integer constants and the symbols
$+, \cdot, \leq, =$) \cite{mcmillan_interpolating_prover}.
For instance, there is no quantifier-free interpolant for the
\laint-formulas $A\defas (2x-y+1=0)$ and $B\defas (y-2z=0)$. 

In order to overcome this problem, different research directions have
been explored. 
One is to restrict to important \emph{fragments} of \laint where
the problem does not occur. To this extent, 
efficient interpolation algorithms for the Difference Logic (\dl) and
Unit-Two-Variables-Per-Inequality (\utvpi) fragments of \laint have
been proposed in \cite{tocl_interpolation}.
Another direction is to extend the signature of \laint to contain
\emph{modular equalities} $=_c$ (or, equivalently,
\emph{divisibility predicates}), so that it is possible to compute
quantifier-free \laint interpolants by means of quantifier elimination
---which is however prohibitively expensive in general, both in theory
and in practice.  For instance, $I\defas (-y+1 =_2 0)\equiv
\exists x.(2x -y+1=0)$ is an interpolant for the formulas $(A,B)$
above.
Using modular equalities, Jain et al. \cite{jain_cav08} developed
polynomial-time interpolation algorithms for linear equations and
their negation and for linear modular equations.  A similar algorithm
was also proposed in \cite{lynch_interpolation}.
The work in \cite{ijcar10_interpolation} was the first to present 
an interpolation algorithm for the full \laint 
(augmented with divisibility predicates) which was not based on
quantifier elimination. 
Finally, an alternative algorithm, 
exploiting efficient interpolation procedures for \larat and for
linear equations in \laint, 
has been %
presented in \cite{lpar10_interpolation}.

The obvious limitation of the first research direction is that it does
not cover the full \laint.  
For the second direction, the approaches so far seem to suffer from 
some drawbacks.
In particular,
some of the interpolation rules of \cite{ijcar10_interpolation}
might result in an exponential blow-up in the size of the interpolants
wrt. the size of the proofs of unsatisfiability from which they are generated.
The algorithm of \cite{lpar10_interpolation} avoids this, 
but at the cost of significantly restricting the heuristics commonly used 
in state-of-the-art \smt solvers for \laint
(e.g. in the framework of \cite{lpar10_interpolation}
both the use of Gomory cuts \cite{ilp_book} 
and of ``cuts from proofs'' \cite{cav09_lia} is not allowed).
More in general, the important issue of how to efficiently integrate 
the presented techniques into a state-of-the-art \smtlaint solver is not 
immediate to foresee from the papers.

In this article we try to close this gap. 
 After recalling the necessary background knowledge
 (\sref{sec:background}), we present 
our contribution, which  is twofold. 

First (\sref{sec:interpolation})  
we show how to extend 
the state-of-the art 
 \laint-solver of \mathsat{} \cite{mathsat5} 
in order to implement interpolant generation on top of it
without affecting its
 efficiency. 
To this extent, we combine different algorithms corresponding to the
different submodules of the \laint-solver, so that each of the
submodules requires only minor modifications,
and implement them in \mathsat (\mathsatcongr hereafter).
An extensive empirical evaluation (\sref{sec:expeval}) shows that 
\mathsatcongr outperforms in efficiency all existing interpolant
generators for \laint.  

Second (\sref{sec:itp_ceiling}), we propose a novel and general
interpolation algorithm for \laint,
{independent from the architecture of \mathsat,}
which overcomes the drawbacks of the current approaches. 
The key idea is to extend both the signature and the domain of \laint:
we extend the signature by adding the \emph{ceiling function}
$\lceil \cdot \rceil$ to it,
and the domain by allowing non-variable terms to be non-integers.
This greatly simplifies the interpolation procedure, 
and allows for producing interpolants which are much more compact
than those generated by the algorithm of \cite{ijcar10_interpolation}.
Also this novel technique was easily implemented on top of the \laint-solver of \mathsat
without affecting its efficiency. (We call this implementation \mathsatceil.)
An extensive empirical evaluation (\sref{sec:expeval}) shows that 
\mathsatceil drastically outperforms \mathsatcongr, 
and hence all other existing interpolant generators for \laint, 
for both efficiency and size of the final interpolant. 

Finally, in \sref{sec:related} we report some related work, and in
\sref{sec:concl} we present some conclusions.
We recall that a shorter version of this article appeared at TACAS 2011
  conference \cite{GriggioLeSebastiani_TACAS11}.

\section{Background: \smtlaint}
\label{sec:background}

We first provide the necessary background. 
{%
We will use the following notational conventions:
\begin{iteMize}{$\bullet$}
\item We denote formulas with $A$, $B$, $S$, $I$, $\varphi$, $\Gamma$.

\item Given a formula $\varphi$ partitioned into $A$ and $B$, the variables in
  $\varphi$ are denoted with $x$, $y$, $z$, 
$s$, %
$v$, $x_i$, $y_j$, $z_k$, $s_h$, $v_l$:
  \begin{iteMize}{$-$}
  \item $x_i$ for variables that occur only in $A$ ($A$-local);
  \item $z_k$ for variables that occur only in $B$ ($B$-local);
  \item $y_j$ for variables that occur both in $A$ and in $B$ ($AB$-common);
  \item $v_l$ when we don't want to distinguish them as in the above
    cases. 
  \end{iteMize}
\item We denote integer constants with $a, b, c, d$.

\item We denote terms with $t_1, t_2$. We write $t_1 \cong t_2$ to denote that
  the two terms are syntactically identical, and $t_1 =_c t_2$ to denote
  that they are congruent modulo $c$. 
  With $\varphi_1 \equiv \varphi_2$ we denote the logical equivalence of the two formulas $\varphi_1$ and $\varphi_2$.

\item We write $t \preceq A$ to denote that all the uninterpreted symbols
  occurring in $t$ occur also in $A$.
  In this case, we say that $t$ is $A$-pure.
  Given two formulas $A, B$ such that $t \preceq (A \cup B)$ but $t \not\preceq A$ and $t \not\preceq B$,
  we say that $t$ is $AB$-mixed.
\end{iteMize}

}
{%

\subsection{Generalities}
\label{sec:background_general}

{In this section we provide some background on  \smt
  (\sref{sec:smt}) and
on interpolation in \smt (\sref{sec:interpolation_smt}).}

\subsubsection{Satisfiability Modulo Theory -- \smt{}}
\label{sec:smt}

Our setting is standard first order logic. 
We use the standard notions of theory, satisfiability, validity,
logical consequence. 
A $0$-ary function symbol is
called a \emph{constant}. A \emph{term} is a first-order term built out of
function symbols and variables. 
If $t_1,\ldots,t_n$
are terms and $p$ is a predicate symbol, then $p(t_1,\ldots,t_n)$ is an
\emph{atom}. A \emph{literal} is either an atom or its negation.
A \emph{formula} $\phi$ is built in the usual way out of the universal and
existential quantifiers, Boolean connectives, and atoms. We call a formula
\emph{quantifier-free} if it does not contain quantifiers, and \emph{ground} if
it does not contain free variables.
A \emph{clause} is a disjunction of
literals. A formula is said to be in \emph{conjunctive normal
  form} (CNF) if it is a conjunction of clauses. 
For every non-CNF \T-formula $\varphi$, an equisatisfiable CNF
formula $\psi$ can be generated in polynomial time \cite{tseitin}.

We call {\em Satisfiability Modulo (the) Theory \T}, \smtt{}, the
problem of deciding the satisfiability of quantifier-free 
formulas wrt. a background theory \T.
\footnote{%
The general definition of \smt deals also with quantified formulas. 
Nevertheless, in this article we restrict our interest to quantifier-free formulas.}
Given a theory \T, we write $\phi \Tmodels \psi$ (or simply
$\phi \models \psi$) to denote that the formula $\psi$ is a logical
consequence of $\phi$ in the theory \T.
With $\phi \preceq \psi$ we denote
that all uninterpreted (in \T) symbols of $\phi$ appear in $\psi$. 
If $C$ is a
clause, $C \downarrow B$ is the clause obtained by removing all the literals
whose atoms do not occur in $B$, and $C \setminus B$ that obtained by
removing all the literals whose atoms do occur in $B$.
With a little abuse of notation, 
we might sometimes denote conjunctions of literals 
$l_1 \wedge \ldots \wedge l_n$ as sets $\{l_1, \ldots, l_n\}$ and vice versa. 
If $\eta$ is the set $\{l_1, \ldots, l_n\}$, 
we might write $\neg\eta$ to mean $\neg l_1 \vee \ldots \vee \neg l_n$.

We call \T-solver a procedure that decides the consistency 
of a conjunction of literals in \T. 
If $S$ %
is a set of literals in \T, 
we call \emph{\T-conflict set w.r.t. $S$} any subset $\eta$ of $S$ 
which is inconsistent in \T.
We call $\neg\eta$ a \emph{\T-lemma} (notice that $\neg\eta$ is a \T-valid clause).  

A standard technique for solving the SMT(\T) problem is to integrate
a DPLL-based SAT solver and a \T-solver in a \emph{lazy} manner
(see, e.g., \cite{sat_handbook_lazy_smt} for a detailed description). DPLL
is used as an enumerator of truth assignments for the
propositional abstraction of the input formula. At each step, the
set of \T-{literals} in the current assignment is sent to the \T-solver
to be checked for consistency in \T. If $S$ is inconsistent, the
\T-solver returns a conflict set $\eta$, and the corresponding
\T-lemma $\neg\eta$ is added as a blocking clause in DPLL, and
used to drive the backjumping and learning mechanism. 

\begin{defi}[Resolution proof]\label{def:resolution_proof}
  Given a set of clauses $S \defas \{C_1,\ldots, C_n\}$ and a clause $C$, we
  call a \emph{resolution proof} {of the deduction} $\bigwedge_i C_i
  \Tmodels C$ a DAG ${\mathcal P}$ such that:\hfill
  \begin{enumerate}[(1)]
  \item $C$ is the root of ${\mathcal P}$;
  \item the leaves of ${\mathcal P}$ are either elements of $S$ or \T-lemmas;
  \item each non-leaf node $C'$ has two {premises} %
    $C_{p_1}$ and $C_{p_2}$ such
    that $C_{p_1} \defas p \vee \phi_1$, $C_{p_2} \defas \neg p \vee \phi_2$,
    and $C' \defas \phi_1 \vee \phi_2$. 
    The atom $p$ is called the
    \emph{pivot} of $C_{p_1}$ and $C_{p_2}$.
  \end{enumerate}
  If $C$ is the empty clause (denoted with $\bot$), then ${\mathcal P}$ is a
  \emph{resolution proof of {(\T-)}unsatisfiability} for $\bigwedge_i C_i$.
\end{defi}

\subsubsection{Interpolation in SMT}
\label{sec:interpolation_smt}

We consider the \smtt{} problem for some background theory \T.
Given an ordered pair $(A, B)$ of formulas such that $A
\wedge B \models_{\T} 
\bot$, a \emph{Craig interpolant} (simply ``interpolant'' hereafter)
is a formula $I$ s.t.
  \begin{enumerate}[(i)]
  \item $A \Tmodels I$,
  \item $I \land B$ is \T-inconsistent, and 
  \item $I \preceq A$ and $I \preceq B$.
  \end{enumerate}
Following \cite{mcmillan_interpolating_prover}, an
interpolant for $(A, B)$ in \smtt 
can be generated by combining a propositional interpolation algorithm 
for the Boolean structure of the formula $A \land B$ 
with a \T-specific interpolation procedure that deals only 
with negations of \T-lemmas (that is, 
with \T-inconsistent conjunctions of \T-literals), as described in Algorithm \ref{alg:SMT_general_schema}. The algorithm works by computing a formula $I_C$ for each clause 
in the resolution refutation, 
such that the formula $I_\bot$ associated to the empty root clause 
is the computed interpolant.
Therefore, in the rest of the article, we shall consider algorithms 
for conjunctions/sets of literals only, 
which can be extended to general formulas by simply ``plugging'' them 
into Algorithm~\ref{alg:SMT_general_schema}.
\begin{minipage}{1.0\linewidth}
\begin{algo}\label{alg:SMT_general_schema}
 \vspace{3mm} \hrule \vspace{2mm} \textbf{Interpolant generation
    for \smtt} 
  \vspace{1mm} \hrule
  \vspace{2mm}\hfill
  \begin{enumerate}[(1)]
  \item Generate a {resolution} proof of unsatisfiability ${\mathcal P}$
    for $A \wedge B$.
  \item For every \T-lemma $\neg\eta$ occurring in ${\mathcal P}$, generate an
    interpolant $I_{\neg\eta}$ for \mbox{$(\eta \setminus B, \eta \downarrow
      B)$}.
  \item For every input clause $C$ in ${\mathcal P}$, set $I_C \defas C \downarrow
    B$ if $C \in A$, and $I_C \defas \top$ if $C \in B$.
  \item For every inner node $C$ of ${\mathcal P}$ obtained by resolution from
    $C_1 \defas {p} \vee \phi_1$ and $C_2 \defas \neg {p} \vee \phi_2$, set
    $I_C \defas I_{C_1} \vee I_{C_2}$ if ${p}$ does not occur in $B$, and $I_C
    \defas I_{C_1} \wedge I_{C_2}$ otherwise.
  \item Output $I_\bot$ as an interpolant for $(A, B)$.
  \end{enumerate}
  \vspace{1mm}
  \hrule \vspace{3mm}
\end{algo}
\end{minipage}

\begin{figure}[t]
  \centering
  \begin{tabular}{c|c}
    \begin{minipage}{0.4\textwidth}
    \newcommand{\TlemmaOne}{$\mathbf{\neg(-y +3x-1 \leq 0) \lor \neg(
        -y-x\leq 0) \lor}$}
    \newcommand{\TlemmaOneb}{$\mathbf{\neg(-z +2y+3\leq 0) \lor \neg(2z-1\leq 0)}$}
    \newcommand{\HypOne}{$\neg(-z +2y +3 \leq 0) \lor (2z-1\leq 0)$}
    \newcommand{\HypTwo}{$p \lor (-z +2y +3 \leq 0)$}
    \newcommand{\HypThree}{$\underline{p \lor (-y +3x-1 \leq 0)}$}
    \newcommand{\HypFour}{$\neg p \lor q$} 
    \newcommand{\HypFive}{$\underline{\neg q \lor
      \neg(-y-x\leq 0)}$} 
    \newcommand{\HypSix}{$\underline{(-y-x\leq 0)}$}

    \newcommand{\ResOne}{$\neg(-y +3x-1 \leq 0) \lor \neg(-y-x\leq 0)
      \lor$} 
    \newcommand{\ResOneb}{$\neg(-z +2y +3 \leq 0)$} 
    \newcommand{\ResTwo}{$\neg(-y +3x-1 \leq 0) \lor \neg(-y-x\leq 0) \lor p$}
    \newcommand{\ResThree}{$\neg(-y-x\leq 0) \lor p$}
    \newcommand{\ResFour}{$\neg(-y-x\leq 0) \lor q$}
    \newcommand{\ResFive}{$\neg(-y-x\leq 0)$}
    \newcommand{\Bottom}{$\bot$}

    \scalebox{0.55}{\input{itp_example_proof.pstex_t}}
  \end{minipage}\hspace{1.8cm} &
  \hspace{1mm}
  \begin{minipage}{0.5\textwidth}
    \newcommand{\TlemmaOne}{$\mathbf{(-4y - 1 \leq 0)}$}
    \newcommand{\HypOne}{$\top$}
    \newcommand{\HypTwo}{$\top$}
    \newcommand{\HypThree}{$p$}
    \newcommand{\HypFour}{$\top$} 
    \newcommand{\HypFive}{$\neg q$}
    \newcommand{\HypSix}{$\bot$}

    \newcommand{\ResOne}{$(-4y - 1 \leq 0)$}
    \newcommand{\ResTwo}{$(-4y - 1 \leq 0)$}
    \newcommand{\ResThree}{$p \lor (-4y - 1 \leq 0)$}
    \newcommand{\ResFour}{$p \lor (-4y - 1 \leq 0)$}
    \newcommand{\ResFive}{$(p \lor (-4y - 1 \leq 0)) \land \neg q$}
    \newcommand{\Bottom}{$(p \lor (-4y - 1 \leq 0)) \land \neg q$}

    \scalebox{0.55}{\input{itp_example_proof2.pstex_t}}
  \end{minipage} \\[3mm]
  \textbf{(a)} & \textbf{(b)}
\end{tabular}
  \caption{Resolution proof of unsatisfiability (a) and interpolant (b) for
    the pair $(A, B)$ of formulas of Example~\ref{ex:itp_example_proof}. In
    the tree on the left, \T-lemmas are displayed in boldface, and clauses
    from $A$ are underlined.
    \label{fig:itp_example_proof}}
\end{figure}
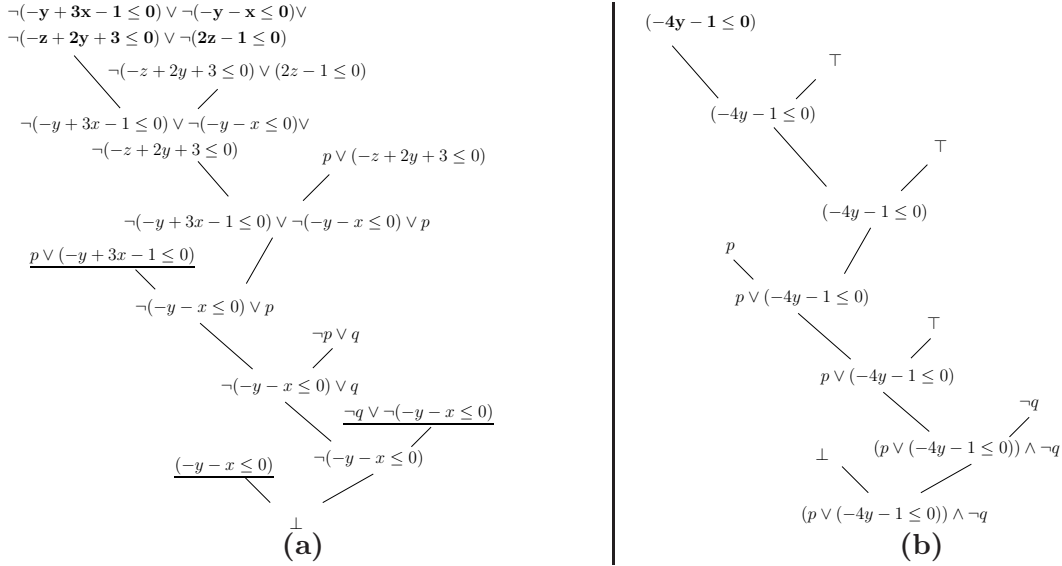

\begin{exa}\label{ex:itp_example_proof}
  {Consider the following two formulas in \larat:}
  \begin{displaymath}
    \begin{split}
      A &\defas (p \lor (-y+3x-1\leq 0)) \land (-y-x\leq 0)
      \land (\neg q \lor \neg(-y-x\leq 0)) \\
      B &\defas (\neg(-z+2y+3\leq 0) \lor (2z-1\leq 0)) \land
      (\neg p \lor q) \land (p \lor (-z+2y+3\leq 0))
    \end{split}
  \end{displaymath}

  Figure~\ref{fig:itp_example_proof}(a) shows a resolution proof of
  unsatisfiability for $A \land B$, in which the clauses from $A$ have been
  underlined. The proof contains the following \larat-lemma (displayed in
  boldface):
  \begin{displaymath}
    \neg(-y +3x-1\leq 0) \lor \neg(-y-x\leq 0) \lor \neg(-z +2y+3\leq 0) \lor \neg(2z-1\leq 0).
  \end{displaymath}

  Figure~\ref{fig:itp_example_proof}(b) shows, for each clause $\Theta_i$ in
  the proof, the formula $I_{\Theta_i}$ generated by Algorithm \ref{alg:SMT_general_schema}. 
  For the
  \larat-lemma, it is easy to see that $(-4y-1\leq 0)$ is an interpolant for
  $((-y +3x-1\leq 0) \land (-y-x\leq 0), (-z +2y+3 \leq 0) \land (2z-1\leq 0))$ as required by Step 2 of the algorithm.
  Therefore, $I_\bot\defas (p \vee (-4y-1\leq 0)) \land
     \neg q$ is an interpolant for $(A, B)$.
\end{exa}
}

\subsection{Efficient \smtlaint solving}
\label{sec:background_smtsolving}

In this section, 
we describe our algorithm for efficiently solving \smtlaint problems,
as implemented in the \mathsat 5 SMT solver \cite{mathsat5}.
They key feature of our solver is an extensive use of \emph{layering}
and \emph{heuristics} for combining different known techniques, in
order to exploit the strengths and to overcome the limitations of each
of them.
{%
Both the experimental results of \cite{mathsat5} 
and the  SMT solvers competition SMT-COMP'10
\footnote{\url{http://www.smtcomp.org/2010/}}
demonstrate that this is a state-of-the-art solver
in \smt$(\laint)$.
}

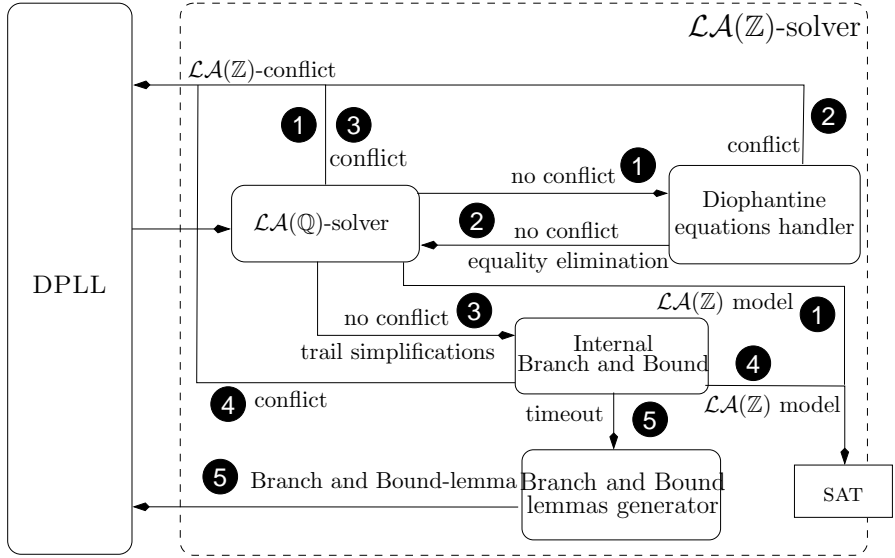
\begin{figure}[t]
  \centering
  \scalebox{0.8}{\input{laint_architecture.pstex_t}}
  \caption{Architecture of the \laint-solver of \mathsat.
    \label{fig:laint_solver_architecture}}
\end{figure}

The architecture of the solver is outlined in Fig.~\ref{fig:laint_solver_architecture}.
It is organized as a layered hierarchy of submodules, 
with cheaper (but less powerful) ones invoked earlier and more often.
The general strategy used for checking the consistency of a set of \laint-constraints is as follows.

First, the rational relaxation of the problem is checked, 
using a Simplex-based \larat-solver similar to that described in \cite{yices}.
If no conflict is detected,
the model returned by the \larat-solver is examined
to check whether all integer variables are assigned to an integer value.
If this happens, the \larat-model is also a \laint-model,
and the solver can return \satres.

Otherwise,
the specialized module for handling linear \laint equations (Diophantine equations) is invoked.
This module is similar to the first part of the Omega test described in \cite{omega}: 
it takes all the equations in the input problem,
and tries to eliminate them by computing a %
solution of the system
and then substituting each variable in the inequalities with its 
expression.
If the system of equations itself is infeasible,
this module is also able to detect the inconsistency, {and to 
produce one  unsatisfiability proof expressed as a linear combination of the
input equations (see \cite{mathsat5} for details).}
Otherwise, 
the inequalities obtained by substituting the variables with their
expressions
are normalized, tightened~%
\footnote{{%
An \laint-inequality 
$\sum_l a_l v_l + c \le 0$ 
can be tightened 
by dividing the constant $c$ by the GCD $g$ of the coefficients,
taking the ceiling of the result, and then multiplying it again by $g$:
$\sum_l a_l v_l + \lceil\frac{c}{g}\rceil\cdot g \le 0$,
s.t. $g\defas GCD(\{a_l\}_l)$.
}
} 
and then sent to the \larat-solver,
in order to check the \larat-consistency of the new set of constraints.

If no conflict is detected, the branch and bound module is invoked,
which tries to find a \laint-solution via branch and bound \cite{ilp_book}.
This module is itself divided into two submodules operating in sequence.
First, the ``internal'' branch and bound module is activated,
which performs case splits directly within the \laint-solver.
The internal search is performed only for a bounded (and small) number of branches, after which the ``external'' branch and bound module is called.
This works in cooperation with the \dpll engine, 
using the ``splitting on-demand'' approach of \cite{splitting_on_demand}:
case splits are delegated to \dpll, by sending to it 
\laint-valid clauses of the form $(t - c \leq 0) \lor (-t + c + 1 \leq 0)$
(called branch-and-bound lemmas)
that encode the required splits.
Such clauses are generated with the ``cuts from proofs'' algorithm of \cite{cav09_lia}:
``normal'' branch-and-bound steps 
-- splitting cases on an individual variable --
are interleaved with ``extended'' steps,
in which branch-and-bound lemmas involve an arbitrary linear combination of variables,
generated by computing proofs of unsatisfiability of particular systems of Diophantine equations.

\section{From \laint-solving to \laint-interpolation}
\label{sec:interpolation}

Our objective is that of devising an interpolation algorithm 
that could be implemented on top of the \laint-solver described in the previous section
without affecting its efficiency.
To this end, we combine different algorithms 
corresponding to the different submodules of the \laint-solver,
so that each of the submodules requires only minor modifications.

\subsection{Interpolation for Diophantine equations}
\label{sec:interpolation_diophantine}
{%
We first consider only conjunctions of positive \laint-equations 
in the form $\sum_l a_lv_l + c = 0$. 
We recall a fundamental property of \laint.
\begin{property}
\label{prop:laint}
The equation  $\sum_l a_lv_l+ c = 0$ is unsatisfiable in \laint if 
the GCD of the coefficients $a_l$ does not divide the constant $c$. 
\end{property}
}

An interpolation procedure for systems of Diophantine equations was given by Jain et al. in \cite{jain_cav08}.
The procedure starts from a proof of unsatisfiability expressed as a 
linear combination of the input equations whose result is an 
\laint-inconsistent equation as in Property~\ref{prop:laint}. 
Given one such proof of unsatisfiability for a system of equations partitioned into $A$ and $B$,
let $(\sum_{x_i \in A \cap B} c_i x_i + \sum_{y_j \not\in B} b_j y_j + c = 0)$
be the linear combination of the equations from $A$ 
with the coefficients given by the proof of unsatisfiability.
Then, {$I \defas \sum_{x_i \in A \cap B} c_i x_i + c =_g 0$,}
where $g$ is any integer that divides $GCD(\{b_j\}_{y_j\not\in B})$, 
is an interpolant for $(A, B)$ \cite{jain_cav08}.

\newcommand{\itp}[1]{\ensuremath{\varphi_{#1}}}
\newcommand{\itpAB}[1]{\ensuremath{\varphi_{#1}^{(A, B)}}}
\newcommand{\itpelem}[2]{\ensuremath{\langle #1, #2 \rangle}}
\newcommand{\aginfer}[2]{\dfrac{#2}{#1}}
{%
\begin{exa}\label{ex:int_equalities}
  Consider the following interpolation problem for the set of equalities
  \begin{displaymath}
  \begin{split}
    &A \defas (-y_1-y_2-4y_3+x_1+2 = 0) \land (-y_3-x_1+x_2=0) \land (-x_1-2x_2+1=0) \\
    &B \defas (7y_1+12y_2+31y_3+10z_1-17=0)
  \end{split}
  \end{displaymath}
One unsatisfiability proof expressed as a linear combination of the
input equations is the following: 

\begin{tiny}
\begin{displaymath}\textstyle
    \infer{
      5y_2-5x_2+10z_1+1= 0 
    }{
      \deduce{4.(-x_1-2x_2+1=0)}{} &
      \infer{
      	  5y_2+4x_1+3x_2+10z_1-3=0
      }{
	    \deduce{3.(-y_3-x_1+x_2=0)}{} &
	    \infer{
	       5y_2+3y_3+7x_1+10z_1-3=0
	    }{
	       \deduce{7.(-y_1-y_2-4y_3+x_1+2=0)}{} &
	       \deduce{7y_1+12y_2+31y_3+10z_1-17=0}{}
	    }
      }  
    }
\end{displaymath}
\end{tiny}

\noindent
{By property~\ref{prop:laint} the root equation $5y_2-5x_2+10z_1+1= 0$ is \laint-inconsistent 
since $GCD(\{5,5,10\})=5$ does not divide $1$.
The proof combines three equations from $A$ with coefficients 7, 3 and
4 respectively. Considering only these equations~
\footnote{or, alternatively, substituting all equations in B with 
the ``true'' equation $0=0$.} we have:}

\begin{tiny}
\begin{displaymath}\textstyle
    \infer{
      -7y_1-7y_2-31y_3-5x_2+18= 0 
    }{
      \deduce{4.(-x_1-2x_2+1=0)}{} &
      \infer{
      	  -7y_1-7y_2-31y_3+4x_1+3x_2+14=0
      }{
	    \deduce{3.(-y_3-x_1+x_2=0)}{} &
	    \infer{-7y_1-7y_2-28y_3+7x_1+14=0
	    }{
	       \deduce{7.(-y_1-y_2-4y_3+x_1+2=0)}{} &
	       \deduce{}{}
	    }
      }  
    }
\end{displaymath}
\end{tiny}

\noindent
Then, {$I \defas -7y_1-7y_2-31y_3+18 =_5 0$,}
is an interpolant for $(A, B)$.
\end{exa}
}

Jain et al. show that a proof of unsatisfiability can be obtained
by computing the Hermite Normal Form \cite{ilp_book} of the system of equations.
However, this is only one possible way of obtaining such proof.
In particular, as shown in \cite{mathsat5}, the submodule of our \laint-solver
that deals with Diophantine equations can  directly produce proofs of unsatisfiability expressed as a 
linear combination of the input equations.
Therefore, we can apply the interpolation algorithm of \cite{jain_cav08} without any modification to the solver.

\subsection{Interpolation for inequalities}
\label{sec:interpolation_congr}

The second submodule of our \laint-solver checks the \larat-consistency 
of a set of inequalities, 
some of which 
obtained by \emph{substitution} and \emph{tightening} \cite{mathsat5}.
In this case, we produce interpolants starting from proofs of unsatisfiability 
in the \emph{cutting-plane proof system}, a complete proof system for
\laint, 
{which is based on 
 the following rules  \cite{ilp_book}:
  \begin{enumerate}[(1)]
  \item{\bf Hyp}
    \begin{math}
      \aginfer{(t \leq 0)}{} 
      \text{~~~~if~ $(t \leq 0)$ is in the input set of \laint-atoms}
    \end{math}
  \vspace{1em}
  \item{\bf Comb} 
    \begin{math}
      \aginfer{
        (c_1t_1 + c_2t_2 \leq 0) 
      }{
        (t_1 \leq 0) \quad 
        (t_2 \leq 0)
      }
    \end{math}
    where:  $c_1, c_2 > 0$
  \vspace{1em}
  \item{\bf Strengthen} 
    \begin{math}
      \aginfer{
        (\sum_i c_i v_i + d \left\lceil \dfrac{c}{d} \right\rceil \leq 0) 
      }{
        (\sum_i c_i v_i + c \leq 0) 
      }
    \end{math}
      where $d > 0$ is an integer that divides all the $c_i$'s.
\end{enumerate}
(Notationally, hereafter we omit representing the {\bf Hyp} rule explicitly,
writing its implied atom as a leaf node in a proof tree; moreover, we
often omit the labels ``{\bf Comb}''.)  }

\subsubsection{Generating cutting-plane proofs in the \laint-solver}

The equality elimination and tightening step generates new inequalities 
$(t' + c' + k \leq 0)$ 
starting from a set of input equalities $\{e_1=0, \ldots, e_n=0\}$ 
and an input inequality $(t + c \leq 0)$.
Thanks to its proof-production capabilities \cite{mathsat5}, 
we can extract from the Diophantine equations submodule the coefficients
$\{ c_1, \ldots, c_n \}$ such that 
$(\sum_i c_i e_i + t + c \leq 0) \equiv (t' + c' \leq 0)$.
Thus, we can generate a proof of $(t' + c' \leq 0)$ by using the Comb and Hyp rules.
We then use the Strengthen rule to obtain a proof of $(t' + c' + k \leq 0)$.
The new inequalities generated are then added to the \larat-solver.
If a \larat-conflict is found, then, the \larat-solver produces 
a \larat-proof of unsatisfiability (as described in \cite{tocl_interpolation})
in which some of the leaves are the new inequalities 
generated by equality elimination and tightening.
We can then simply replace such leaves with the corresponding cutting-plane proofs to 
obtain the desired cutting-plane unsatisfiability proof.

{%
\begin{exa}\label{ex:equa_elimination_tightening}
  Consider the following sets of \laint-constraints: 
    \begin{displaymath}
      E \defas \left\{
        \begin{array}{l}
          2v_1- 5v_3 = 0 \\
          v_2- 3v_4 = 0
        \end{array}
      \right. \quad\quad\quad
      I \defas \left\{
        \begin{array}{l}
           - 2v_1- v_2- v_3 + 7\leq 0 \\
          2v_1+ v_2+ v_3 - 8\leq 0
        \end{array}
      \right.
    \end{displaymath}
$E \cup I$ is satisfiable over the rationals, but not over the integers.
Therefore, the \laint-solver invokes the equality elimination procedure,
which generates a new set $I'$ of inequalities by ``inlining'' the equalities of $E$ in $I$.
In particular, $I'$ is generated as follows:

\begin{small}
  \begin{equation}\label{example-proof-eq-elim}
    \begin{array}{rcl}
      -5 \cdot (2v_1- 5v_3 = 0), 1 \cdot (v_2- 3v_4 = 0), ( -2v_1- v_2- v_3 + 7\leq 0) &\!\!\!\leadsto\!\!\! & (-3v_4  -  12 v_1+ 24 v_3 + 7 \leq 0) \\
      5 \cdot (2v_1- 5v_3 = 0), -1 \cdot (v_2- 3v_4 = 0), (2v_1+ v_2+ v_3 - 8\leq 0) &\!\!\!\leadsto\!\!\! & (3v_4 + 12v_1- 24 v_3 - 8 \leq 0)
    \end{array}
  \end{equation}
\end{small}

The inequalities in $I'$ can now be tightened by dividing the constant by the GCD of the coefficients,
taking the ceiling of the result, and then multiplying again: 

\begin{displaymath}
  I'' = \left\{
    \begin{array}{lcl}
      -3v_4 - 12 v_1+ 24 v_3 + \left\lceil\dfrac{7}{3}\right\rceil\cdot 3 \leq 0 & \text{~which becomes~} & -3v_4 - 12v_1+ 24v_3 + 9 \leq 0 \\
      3v_4 + 12v_1- 24 v_3 + \left\lceil\dfrac{- 8}{3}\right\rceil\cdot 3 \leq 0 & \text{~which becomes~} & 3v_4  + 12v_1- 24v_3 - 6 \leq 0 \\
    \end{array}
  \right.
\end{displaymath}

$I''$ is then sent back to the \larat-solver, which can now easily detect its inconsistency,
producing the following \larat-proof of unsatisfiability $P_{\larat}$ for it:

\begin{displaymath}
  P_{\larat} \defas \vcenter{\infer{3 \leq 0}{
    1 \cdot (-3v_4 - 12v_1+ 24v_3 + 9 \leq 0) &
    1 \cdot (3v_4 + 12v_1 - 24v_3 - 6 \leq 0)
  }}
\end{displaymath}

The final cutting-plane proof $P_{\laint}$ for the \laint-unsatisfiability of $E \cup I$ can then
be constructed by replacing the two inequalities in $P_{\larat}$ 
with their proofs $P_1$ and $P_2$ constructed with the information \eqref{example-proof-eq-elim} 
computed by the equality elimination procedure:

\begin{displaymath}
  P_1 \defas \vcenter{
  \infer[\text{[Strengthen]}]{
    -3v_4 - 12v_1+ 24v_3 + 9 \leq 0
  }{
    \infer[\text{}]{-3v_4 - 12v_1+ 24v_3 + 7 \leq 0}{
      \deduce{5 \cdot (-2v_1+ 5v_3\leq 0)}{} &
      \infer[\text{}]{
        -3v_4 - 2v_1- v_3 + 7\leq 0
      }{
        v_2- 3v_4\leq 0 &
        -2v_1- v_2- v_3 + 7\leq 0
      }
    }
  }}
\end{displaymath}

\begin{displaymath}
  P_2 \defas \vcenter{
  \infer[\text{[Strengthen]}]{3v_4 + 12v_1- 24v_3 - 6 \leq 0}{
    \infer[\text{}]{
      3v_4 + 12v_1- 24v_3 - 8 \leq 0
    }{
      \deduce{5 \cdot (2v_1- 5v_3\leq 0)}{} &
      \infer[\text{}]{
        3v_4 + 2v_1+ v_3 - 8\leq 0
      }{
        -v_2+ 3v_4\leq 0 &
        2v_1+ v_2+ v_3 - 8\leq 0
      }
    } 
  }}
\end{displaymath}

\begin{displaymath}
  P_{\laint} \defas \vcenter{
  \infer{
    3 \leq 0
  }{
    \infer{1 \cdot (-3v_4 - 12v_1+ 24v_3 + 9 \leq 0)}{P_1} &
    \infer{1 \cdot (3v_4 + 12v_1- 24v_3 - 6 \leq 0)}{P_2}
  }}.
\end{displaymath}

\end{exa}
}

\subsubsection{From proofs to interpolants.}

In analogy to previous work on \larat and \laint \cite{mcmillan_interpolating_prover,ijcar10_interpolation},
we produce interpolants by annotating each step of the proof of unsatisfiability of $A \land B$,
such that the annotation for the root of the proof
(deriving an inconsistent inequality $(c \leq 0)$ with $c \in \mathbb{Z}^{>0}$)
is an interpolant for $(A, B)$.

\begin{defi}[Valid annotated sequent]
  An \emph{annotated sequent} is a sequent in the form
\label{eq:annotated_sequent}
$    
(A, B) \vdash (t \leq 0)[I] 
$
  where $A$ and $B$ are conjunctions of equalities and inequalities in \laint,
  and where $I$ (called \emph{annotation}) is a set of pairs
  $\itpelem{(t_i \leq 0)}{E_i}$ 
  in which $E_i$ is a (possibly empty) conjunction of equalities and modular equalities.
  It
  is said to be \emph{valid} when:
  \begin{enumerate}[(1)]
  \item $A \models \bigvee_{\itpelem{t_i \leq 0}{E_i} \in I} 
                           ((t_i \leq 0) \land E_i)$;
  \item For all $\itpelem{t_i \leq 0}{E_i} \in I$, 
    $B \land E_i \models (t - t_i \leq 0)$;
  \item For every element $\itpelem{(t_i \leq 0)}{E_i}$ of $I$,
    $t_i \preceq A$, $(t - t_i) \preceq B$, $E_i \preceq A$ and $E_i \preceq B$.
  \end{enumerate}
\end{defi}

\begin{defi}[Interpolating Rules]
  The \laint-interpolating inference rules that we use are the following:\hfill
  \begin{enumerate}[(1)]
  \item{\bf Hyp-A}
    \begin{math}
      \aginfer{(A, B) \vdash (t \leq 0) 
                           [\{\itpelem{t \leq 0}{\top} \}]}{} 
      \text{~~~~if~} (t \leq 0) \in A \text{~or~} (t = 0) \in A
    \end{math}

  \vspace{1ex}
  \item{\bf Hyp-B} 
    \begin{math}
      \aginfer{(A, B) \vdash (t \leq 0) 
                           [\{\itpelem{0 \leq 0}{\top} \}]}{} 
      \text{~~~~if~} (t \leq 0) \in B \text{~or~} (t = 0) \in B
    \end{math}

  \vspace{1em}
  \item{\bf Comb} 
    \begin{math}
      \aginfer{
        (A, B) \vdash (c_1t_1 + c_2t_2 \leq 0) [I]
      }{
        (A, B) \vdash (t_1 \leq 0) [I_1] \quad 
        (A, B) \vdash (t_2 \leq 0) [I_2]
      }
    \end{math}
    where:
    \begin{iteMize}{$-$}
    \item $c_1, c_2 > 0$
    \item $I \defas \{\itpelem{c_1t'_1 + c_2t'_2 \leq 0}{E_1 \land E_2} ~|~
                 \itpelem{t'_1 \leq 0}{E_1} \in I_1 \text{~and~}
                 \itpelem{t'_2 \leq 0}{E_2} \in I_2 \}$
    \end{iteMize}

  \vspace{1em}
  \item{\bf Strengthen} 
    \begin{math}
      \aginfer{
        (A, B) \vdash (\sum_i c_i x_i + c + k \leq 0) [I]
      }{
        (A, B) \vdash (\sum_i c_i x_i + c \leq 0) 
                      [\{\itpelem{t' \leq 0}{\top}\}]
      }
    \end{math}
    where:
    \begin{iteMize}{$-$}
    \item $k \defas d \left\lceil \dfrac{c}{d} \right\rceil - c$, and
    $d > 0$ is an integer that divides all the $c_i$'s;
    \item $I \defas \{ \itpelem{t' + j \leq 0}
                               {\exists(x \not\in B).(t' + j = 0)} ~|~ 
                       0 \leq j < k \} \cup 
                    \{ \itpelem{t' + k \leq 0}{\top} \}$; and
    \item $\exists(x \not\in B).(t' + j = 0)$ denotes 
      the result of the existential elimination from $(t' + j = 0)$ of 
      all and only the variables $x_1,...,x_n$ not occurring in $B$.%
    \end{iteMize}
{(We recall that 
        $\exists(x_1,\ldots,x_n).(\sum_i c_i x_i + \sum_j d_j y_j + c = 0) \equiv (\sum_j d_j y_j + c =_{GCD(c_i)} 0)$, 
        and that $(t =_0 0) \equiv (t = 0)$.)}
  \end{enumerate}
\end{defi}

\begin{thm}
\label{teo:inequalities_int}
  All the interpolating rules preserve the validity of the sequents.
\end{thm}
{%
\proof
  In the following, let 
  \begin{math}
      \itp{I} \defas %
          \bigvee_{\itpelem{t_i \leq 0}{E_i} \in I} 
                          ((t_i \leq 0) \land E_i). %
  \end{math}\hfill
  \begin{enumerate}[(1)]
   \vspace{1ex}
  \item {\bf Hyp-A}: obvious.

  \vspace{1em}
  \item {\bf Hyp-B}: obvious.

  \vspace{1em}
  \item {\bf Comb}
    \begin{enumerate}[(3.1)]
    \item By hypothesis, we have $A \models \itp{I_1}$ and 
      $A \models \itp{I_2}$. Therefore
      \begin{displaymath}
        A \models (\bigvee_{I_1} (t'_{1i} \leq 0 \land E_{1i})) \land
                  (\bigvee_{I_2} (t'_{2j} \leq 0 \land E_{2j})).
      \end{displaymath}
      By applying DeMorgan's rules:
      \begin{displaymath}
        \begin{split}
          A \models  & \bigvee_{I_1} (
                     (t'_{1i} \leq 0 \land E_{1i}) \land
                     (\bigvee_{I_2} (t'_{2j} \leq 0 \land E_{2j}))) 
                     \equiv \\
                     & \bigvee_{I_1} 
                       \bigvee_{I_2} (
                       \underbrace{(t'_{1i} \leq 0 \land t'_{2j} \leq 0)}_{\psi_{ij}}
                       \land E_{1i} \land E_{2j}).
        \end{split}
      \end{displaymath}
      Now, since $c_1, c_2 > 0$, we have that 
      $\psi_{ij} \models (c_1 t'_{1i} + c_2 t'_{2j} \leq 0)$; therefore
      \begin{displaymath}
        A \models \bigvee_{I_1} 
                  \bigvee_{I_2} (
                       (c_1 t'_{1i} + c_2 t'_{2j} \leq 0)
                       \land E_{1i} \land E_{2j}) \equiv 
                  \itp{I}.
      \end{displaymath}

    \item By hypothesis, we have
      \begin{displaymath}
        \begin{split}
          & B \land E_{1i} \models (t_1 - t'_{1i} \leq 0) \text{~and} \\
          & B \land E_{2j} \models (t_2 - t'_{2j} \leq 0)
        \end{split}
      \end{displaymath}
      for all $\itpelem{t'_{1i}}{E_{1i}} \in I_1$ and 
              $\itpelem{t'_{2j}}{E_{2j}} \in I_2$.
      Therefore:
      \begin{displaymath}
        \begin{split}
        B \land E_{1i} \land E_{2j} \models~ &
        (t_1 - t'_{1i} \leq 0) \land (t_2 - t'_{2j} \leq 0) \\
        \models~ &
        ((c_1t_1 - c_1t'_{1i}) + (c_2t_2 - c_2t'_{2j}) \leq 0) 
        \equiv \\
        & ((c_1t_1 + c_2t_2) - (c_1t'_{1i} - c_2t'_{2j}) \leq 0).
        \end{split}
      \end{displaymath}

    \item Follows immediately from the hypothesis.
    \end{enumerate}

    \vspace{1em}
  \item{\bf Strengthen}
    \begin{enumerate}[(4.1)]
    \item We observe that in this case $\itp{I}$ is equivalent to 
      \begin{displaymath}
        \itp{I}' \defas \bigvee_{0 \leq j < k} \exists(x\not\in B).(t' + j = 0) \lor (t' + k \leq 0).
      \end{displaymath}
      We also observe that, in \laint, 
      $(t' \leq 0)$ is equivalent to
      \footnote{In fact, this is true 
        for all $t'$ and all $k \in \mathbb{Z}^{\geq 0}$.}
      \begin{displaymath}
        \psi_I \defas \bigvee_{0 \leq j < k} (t' + j = 0) \lor (t' + k \leq 0).
      \end{displaymath}
      By hypothesis, $A \models (t' \leq 0)$, and thus $A \models \psi_I$.
      Since $\psi_I \models \itp{I}'$, 
      we can immediately conclude.
    \item The hypothesis in this case is:
      \begin{displaymath}
        B \models ((\sum_i c_i x_i + c) - t' \leq 0).
      \end{displaymath}
      We want to prove that
      \begin{enumerate}
      \item $B \land \exists(x \not\in B).(t' + j = 0) \models ((\sum_i c_i x_i + c + k) - (t' + j) \leq 0)$
        for all $0 \leq j < k$; and
      \item $B \models ((\sum_i c_i x_i + c + k) - (t' + k) \leq 0)$.
      \end{enumerate}
      The latter follows immediately from the hypothesis.

      As regards (i), from the hypothesis and the fact that 
      $(t' + j = 0) \models (t' + j \leq 0)$ we have 
      \begin{displaymath}
       \qquad \qquad B \land (t' + j = 0) \models 
        (((\sum_i c_i x_i + c) - t') + (t' + j) \leq 0) \equiv
        (\sum_i c_i x_i + c + j \leq 0).
      \end{displaymath}
      But then
      \begin{equation}\label{eq:strengthen-valid-b}
        B \land (t' + j = 0) \models 
        (\sum_i c_i x_i + d\left\lceil\dfrac{c + j}{d}\right\rceil \leq 0).
      \end{equation}
      where $d > 0$ divides all the $c_i$'s.
      By definition, $k = d\left\lceil\dfrac{c}{d}\right\rceil - c$ and $j < k$.
      Therefore $\dfrac{c + j}{d} < \left\lceil\dfrac{c}{d}\right\rceil$ (since $d > 0$),
      and thus $\left\lceil\dfrac{c + j}{d}\right\rceil \leq \left\lceil\dfrac{c}{d}\right\rceil$.
      But since $j \geq 0$, $\dfrac{c + j}{d} \geq \dfrac{c}{d}$, and so
      it must be $\left\lceil\dfrac{c + j}{d}\right\rceil = \left\lceil\dfrac{c}{d}\right\rceil$.
      From this fact and \eqref{eq:strengthen-valid-b} it follows that
      \begin{displaymath}
        B \land (t' + j = 0) \models 
        (\sum_i c_i x_i + d\left\lceil\dfrac{c}{d}\right\rceil \leq 0) \equiv        
        (\sum_i c_i x_i + c + k \leq 0).
      \end{displaymath}
      Since $(t' + j = 0) \models (-(t' + j) \leq 0)$, then
      \begin{displaymath}
      \begin{split}
        B \land (t' + j = 0) \models\, &
        (\sum_i c_i x_i + c + k \leq 0) \land (-(t' + j) \leq 0) \\ \models\, &
        (\sum_i c_i x_i + c + k -(t' + j) \leq 0).
      \end{split}
      \end{displaymath}
      Therefore,
      \begin{displaymath}
        \exists(x\not\in B).(B \land (t' + j = 0)) \models
        \exists(x\not\in B).(\sum_i c_i x_i + c + k -(t' + j) \leq 0).
      \end{displaymath}
      We can then conclude by observing that:
      \begin{iteMize}{$-$}
      \item Trivially, 
        $\exists(x\not\in B)(B\land (t'+j=0)) \equiv B\land\exists(x\not\in B).(t'+j=0)$; and
      \item From the validity of the premise of the Strengthen rule, we have that 
        $(\sum_i c_i x_i + c -t'\leq 0) \preceq B$, and thus
        $\exists(x\not\in B).(\sum_i c_i x_i + c + k -(t' + j) \leq 0) \equiv (\sum_i c_i x_i + c + k -(t' + j) \leq 0)$.\\
      \end{iteMize}
    \item Follows immediately from the hypothesis and the fact that 
      variables not occurring in $B$ are eliminated from equations.
      \qed
    \end{enumerate}
  \end{enumerate}
}
\begin{cor}
\label{cor:inequalities_int}
  If we can derive a valid sequent $(A, B) \vdash c \leq 0 [I]$ with 
  $c \in \mathbb{Z}^{> 0}$, then %
  $%
      \itp{I} \defas  \bigvee_{\itpelem{t_i \leq 0}{E_i} \in I} 
                       ((t_i \leq 0) \land E_i)
  $
is an interpolant for $(A, B)$.
\end{cor}
{%
\proof\hfill
  \begin{enumerate}[(1)]
  \item $\mathbf{A\models \itp{I}.}$ Trivial from the first validity condition.
  \item $\mathbf{B\land \itp{I}\models \bot.}$ From the second validity condition, we have
    \begin{displaymath}
      B \land E_i \models (c - t_i \leq 0).
    \end{displaymath}
    for all $\itpelem{t_i \leq 0}{E_i} \in I$.
    Therefore,
    \begin{displaymath}
      B \land E_i \land \neg (c - t_i \leq 0) \models \bot.
    \end{displaymath}
    Since $c \in \mathbb{Z}^{>0}$, 
    $\neg (c - t_i \leq 0)$ is entailed in \laint by $(t_i \leq 0)$, 
    and thus
    \begin{displaymath}
      B \land E_i \land (t_i \leq 0) \models \bot
    \end{displaymath}
    for all $\itpelem{t_i \leq 0}{E_i} \in I$. 
    Thus, $B \land \itp{I} \models \bot$.
  \item $\mathbf{\itp{I}\preceq A}$ and $\mathbf{\itp{I}\preceq B}$.
    Trivial from the third validity condition.
    \qed  
  \end{enumerate}
  \vspace{1ex}  
}
Notice that the first three rules correspond to the rules for \larat given in \cite{mcmillan_interpolating_prover},
whereas Strengthen is a reformulation of the $k$-Strengthen rule given in \cite{ijcar10_interpolation}.
Moreover, although the rules without annotations are refutationally complete for \laint,
in the above formulation the annotation of Strengthen might prevent its applicability,
thus losing completeness.
In particular, it only allows to produce proofs with at most one strengthening per branch.
Such restriction has been put only for simplifying the proofs of correctness,
and it is not present in the original $k$-Strengthen of \cite{ijcar10_interpolation}.
However, for our purposes this is not a problem, 
since we use the above rules only in the second submodule of our \laint-solver,
which always produces proofs with at most one strengthening per branch.
{%
\begin{exa}\label{ex:itp_app}
  Consider the following interpolation problem \cite{lpar10_interpolation}: 
  \begin{displaymath}
  \begin{split}
    &A \defas (-y_1 -10x_1 -4 \leq 0) \land (y_1 + 10x_1 \leq 0) \\
    &B \defas (-y_1 -10z_1 + 1 \leq 0) \land (y_1 + 10z_1 - 5 \leq 0).
  \end{split}
  \end{displaymath}
Using the above interpolating rules, 
we can construct the following annotated cutting-plane proof of unsatisfiability:
\begin{small}
\begin{displaymath}\textstyle
    \infer{
      1 \leq 0 ~
      \begin{array}{l}
      [\{\itpelem{j-4 \leq 0}{\exists x_2.(y_1 + 10x_2 + j = 0)}~|~0 \leq j < 9\} \cup 
      \phantom{[}\{\itpelem{(5 \leq 0}{\top}\}]
       \end{array}
    }{
      \infer{
	\deduce{
          \begin{array}{l}
            [\{\itpelem{y_1 + 10x_1 + j \leq 0}{\\~~~~\exists x_2.(y_1 + 10x_2 + j = 0)}~|~
            ~~~~0 \leq j < 9\} \cup \\
            \{\itpelem{y_1 + 10x_1 + 9 \leq 0}{\top}\}]
          \end{array}
        }{10x_1 - 10z_1 + 10 \leq 0}
      }{
	\infer{
	  \deduce{[\{\itpelem{y_1 + 10x_1 \leq 0}{\top}\}]}{10x_1 - 10z_1 + 1 \leq 0}
	}{
	  \deduce{[\{\itpelem{y_1 + 10x_1 \leq 0}{\top}\}]}{y_1 + 10x_1 \leq 0} &
	  \deduce{[\{\itpelem{0 \leq 0}{\top}\}]}{-y_1 - 10z_1+ 1 \leq 0}
	}
      } &
      \infer{
	\deduce{[\{\itpelem{-y_1 - 10x_1 - 4 \leq 0}{\top}\}]}{-10x_1 + 10z_1 -9 \leq 0}
      }{
	\deduce{[\{\itpelem{-y_1 - 10x_1 - 4 \leq 0}{\top}\}]}{-y_1 - 10x_1 - 4 \leq 0} &
	\deduce{[\{\itpelem{0 \leq 0}{\top}\}]}{y_1 + 10z_1 - 5\leq 0}
      }
    }  
\end{displaymath}
\end{small}
Since $(j-4 \leq 0) \models \bot$ when $j \geq 5$,
the generated interpolant $I$ is:
\begin{small}
\begin{displaymath}
\begin{split}
I 
& \defas \exists x_2.((y_1 + 10x_2=0) \lor (y_1+10x_2+1=0) \lor \ldots \lor (y_1+10x_2+4=0)) \\
& = (y_1 =_{10} 0) \lor (y_1 =_{10} -1) \lor (y_1 =_{10} -2) \lor (y_1=_{10}-3) \lor (y_1=_{10}-4)
\end{split}
\end{displaymath}
\end{small}
\end{exa}

\subsubsection{Conditional strengthening}
In \cite{ijcar10_interpolation} some optimizations of the
$k$-Strengthen rule are given for some special cases.
Here, we present another one, which lets us avoid performing case splits under certain conditions,
and thus results in more concise interpolants than the general {\bf Strengthen} rule.
In particular,
if \emph{both} the result of a Strengthen rule 
  and the linear combination of all the inequalities from $B$ in the 
  subtree on top of the Strengthen contain only $AB$-common symbols,
  then 
it is possible to perform a \emph{conditional strengthening} as follows:
  \begin{description}
  \item[Conditional-Strengthen] 
    \begin{displaymath}
      \infer{
        (A, B) \vdash \sum_i c_i x_i + c + k \leq 0 [I]
      }{
        (A, B) \vdash \sum_i c_i x_i + c \leq 0 
                      [\{\itpelem{t' \leq 0}{\top}\}]
      }
    \end{displaymath}
    where:
    \begin{iteMize}{$-$}
    \item $k \defas d \left\lceil \dfrac{c}{d} \right\rceil - c$;
    \item $d > 0$ is an integer that divides all the $c_i$'s;
    \item $\sum_i c_i x_i + c + k$ is $AB$-common;
    \item $I \defas \{\itpelem{\neg P}{\neg P},
    \itpelem{\sum_i c_i x_i + c +k \leq  0}{\top}\}$;
    \item $P \defas \sum_i c_i x_i + c -t'\leq 0 $ is the single inequality obtained by combining 
      all the constraints from $B$ in the subtree on top of the premise 
      (with the coefficients occurring in the subtree);
    \item $P$ is $AB$-local.
    \end{iteMize}
  \end{description}

  We observe that this is similar to what we do 
  in Case 4 of our interpolation algorithm for \utvpi
  \cite{tocl_interpolation}. 
  Further, notice that in the definition above, with a little abuse of notation,
  we are storing an inequality as the second component of the first pair of $I$,
  and not an equality.
  However, this does not affect the validity of {\bf Hyp-A}, {\bf Hyp-B} and {\bf Comb}. 
  Together with the fact that we only generate proofs with at most one strengthening per branch (see above),
  the following Theorem is then enough to ensure that Corollary~\ref{cor:inequalities_int} still holds.

\begin{thm}
  \label{teo:inequalities_int_optimized}
  {\bf Conditional-Strengthen} preserves the validity of the sequents.
\end{thm}
\proof \hfill
       \begin{enumerate}[(1)]
        \item 
         We observe that in this case $\itp{I}$ is equivalent to $\itp{I}'$ where
         \begin{displaymath}
         \itp{I}' \defas (\neg P \lor (\sum_i c_i x_i + c + k \leq 0)) \equiv (P \rightarrow (\sum_i c_i x_i + c + k \leq 0))
         \end{displaymath}
          By hypothesis, $A \models (t' \leq 0)$, and by definition $P\cong (\sum_i c_i x_i + c - t' \leq 0)$. Therefore, 
          \begin{displaymath}
          \begin{split}
          A \land P &\models (\sum_i c_i x_i +c \leq 0), \text{~and so} \\
          A \land P &\models  (\sum_i c_i x_i + c + k \leq 0). \\
           \end{split}
          \end{displaymath}
          Hence $A \models (\neg P \lor (\sum_i c_i x_i + c + k \leq 0)) \equiv \itp{I}$.

\vspace{1em}
        \item We observe that 
        \begin{displaymath}
        \neg P \cong (-\sum_i c_i x_i - c + t' +1 \leq 0).
        \end{displaymath} 
        By hypothesis, $B \models (\sum_i c_i x_i + c - t' \leq 0) \cong P$, hence $B \land \neg P \models \bot$.
        Therefore, we can conclude that: 
          \begin{enumerate}
          \item $B \land \neg P \models ((\sum_i c_i x_i + c + k) - (-\sum_i c_i x_i - c + t'+1) \leq 0) $, and
          \item $B \models ((\sum_i c_i x_i + c + k) - (\sum_i c_i x_i + c + k) \leq 0)$.
          \end{enumerate}

\vspace{1em}
        \item Follows immediately from the hypothesis.
          \qed
        \end{enumerate}
\begin{exa}\label{ex:itp_conditional_strengthen}
  Consider the following variant of Example~\ref{ex:itp_app}:
  \begin{displaymath}
  \begin{split}
    &A \defas (-y_1 -10y_3 -4 \leq 0) \land (y_1 + 10y_3 \leq 0) \land (y_2+x_1\leq 0)\\
    &B \defas (-y_1 -10y_2 + 1 \leq 0) \land (y_1 + 10y_2 - 5 \leq 0) \land (y_3+z_1\leq 0)
  \end{split}
  \end{displaymath}

\noindent By applying {\bf Hyp-A}, {\bf Hyp-B}, {\bf Comb} and {\bf Strengthen} rules, we can obtain the following unsatisfiability proof with an interpolant $I_1$:

\begin{small}
\begin{displaymath}\textstyle
    \infer{
      1 \leq 0 ~
      \begin{array}{l}
      [I_1]
       \end{array}
    }{
      \infer{
	\deduce{
          \begin{array}{l}
            [\{\itpelem{y_1 + 10y_3 + j \leq 0}{\\~~~~y_1 + 10y_3 + j = 0}~|~
            ~~~~0 \leq j < 9\} \cup \\
            \{\itpelem{y_1 + 10y_3 + 9 \leq 0}{\top}\}]
          \end{array}
        }{10y_3 - 10y_2 + 10 \leq 0}
      }{
	\infer{
	  \deduce{[\{\itpelem{y_1 + 10y_3 \leq 0}{\top}\}]}{10y_3 - 10y_2 + 1 \leq 0}
	}{
	  \deduce{[\{\itpelem{y_1 + 10y_3 \leq 0}{\top}\}]}{y_1 + 10y_3 \leq 0} &
	  \deduce{[\{\itpelem{0 \leq 0}{\top}\}]}{-y_1 - 10y_2 + 1 \leq 0}
	}
      } &
      \infer{
	\deduce{[\{\itpelem{-y_1 - 10y_3 - 4 \leq 0}{\top}\}]}{-10y_3 + 10y_2 -9 \leq 0}
      }{
	\deduce{[\{\itpelem{-y_1 - 10y_3 - 4 \leq 0}{\top}\}]}{-y_1 - 10y_3 - 4 \leq 0} &
	\deduce{[\{\itpelem{0 \leq 0}{\top}\}]}{y_1 + 10y_2 - 5\leq 0}
      }
    }  
\end{displaymath}
\end{small}

\noindent where $I_1$ is defined as follows:

\begin{small}
\begin{displaymath}
\begin{split}
I_1& \defas  \{\itpelem{j-4 \leq 0}{y_1 + 10y_3 + j = 0}~|~0 \leq j < 9\} \cup 
      \phantom{[}\{\itpelem{(5 \leq 0}{\top}\},\\
      &= (y_1 + 10y_3=0) \lor (y_1+10y_3+1=0) \lor (y_1+10y_3+2=0) \lor (y_1+10y_3+3=0) \lor \\ & \phantom{=~~~} (y_1+10y_3+4=0)
\end{split}
\end{displaymath}
\end{small}

We observe from the above proof that all the inequalities from $B$ above the tightened inequality $(10y_3-10y_2+1\leq 0)$, and this inequality itself, contain only $AB$-common symbols. Therefore, we can replace the {\bf Strengthen} rule with the {\bf Conditional-Strengthen} rule in the above proof to obtain a more concise interpolant $I_2$:

\begin{small}
\begin{displaymath}\textstyle
    \infer{
      1 \leq 0 ~
      \begin{array}{l}
      [I_2]
       \end{array}
    }{
      \infer{
	\deduce{
          \begin{array}{l}
            [\{\itpelem{y_1 + 10y_2 \leq 0}{y_1+10y_2 \leq 0}, \\
            \itpelem{10y_3-10y_2+10 \leq 0}{\top}\}]
          \end{array}
        }{10y_3 - 10y_2 + 10 \leq 0}
      }{
	\infer{
	  \deduce{[\{\itpelem{y_1 + 10y_3 \leq 0}{\top}\}]}{10y_3 - 10y_2 + 1 \leq 0}
	}{
	  \deduce{[\{\itpelem{y_1 + 10y_3 \leq 0}{\top}\}]}{y_1 + 10y_3 \leq 0} &
	  \deduce{[\{\itpelem{0 \leq 0}{\top}\}]}{-y_1 - 10y_2 + 1 \leq 0}
	}
      } &
      \infer{
	\deduce{[\{\itpelem{-y_1 - 10y_3 - 4 \leq 0}{\top}\}]}{-10y_3 + 10y_2 -9 \leq 0}
      }{
	\deduce{[\{\itpelem{-y_1 - 10y_3 - 4 \leq 0}{\top}\}]}{-y_1 - 10y_3 - 4 \leq 0} &
	\deduce{[\{\itpelem{0 \leq 0}{\top}\}]}{y_1 + 10y_2 - 5\leq 0}
      }
    }  
\end{displaymath}
\end{small}

\noindent where $I_2$ is defined as follows:

\begin{small}
\begin{displaymath}
\begin{split}
I_2 
&\defas \{\itpelem{10y_2-10y_3-4 \leq 0}{y_1 + 10y_2 \leq 0}, 
      \phantom{[}\itpelem{(-y_1-10y_2+6 \leq 0}{\top}\} \\
& =
      ((10y_2-10y_3-4 \leq 0) \land (y_1 + 10y_2 \leq 0))\lor  
      (-y_1-10y_2+6 \leq 0),
\end{split}
\end{displaymath}
\end{small}

\noindent
(which can then be simplified to 
$((y_2-y_3 \leq 0) \land (y_1 + 10y_2 \leq 0))\lor (-y_1-10y_2+6 \leq 0)$.)
\end{exa}
}

\subsection{Interpolation with branch-and-bound}
\label{sec:interpolation_bb}

\subsubsection{Interpolation via splitting on-demand}
In the splitting on-demand approach, the \laint solver might not
always detect the unsatisfiability of a set of constraints by itself;
rather, it might cooperate with the \dpll solver by asking it to
perform some case splits, by sending to \dpll some additional
\laint-lemmas encoding the different case splits.  In our
interpolation procedure, we must take this possibility into account.

Let 
$(t - c\leq 0) \lor (-t+c+1 \leq 0)$ 
be a branch-and-bound lemma 
added to the \dpll solver by the \laint-solver, using splitting on-demand.
If $t \preceq A$ or $t \preceq B$, then we can exploit the Boolean interpolation algorithm
also for computing interpolants in the presence of splitting-on-demand lemmas.
The key observation is that the lemma
$(t - c\leq 0) \lor (-t+c+1 \leq 0)$ 
is a \emph{valid clause} in \laint.
Therefore, we can add it to any formula without affecting its satisfiability. 
Thus, if $t \preceq A$ we can treat the lemma as a clause from $A$, 
and if $t \preceq B$ we can treat it as a clause from $B$; %
{
 if both $t \preceq A$ and $t \preceq B$, we are free to choose between
the two alternatives.
}

{%
  \begin{exa}
\label{ex:splitting-on-demand}
Consider the following \laint set of constraints $S$, which is first
fed to the \larat-Solver, producing the \larat model $\mu_S$:
  \begin{displaymath}
    S \defas \left\{
      \begin{array}{l}
          y_1 + 5 y_2 - 5 y_3 - 2 x_1 + 2 \leq 0 \\
          - y_1 - 5 y_2 + 5 y_3 + 4 z_1 - 3 \leq 0 \\
          x_1 \leq 0 \\
          y_1 \leq 0 \\
          -y_1 \leq 0 \\
          -y_2 \leq 0 \\
          y_2 - 2 \leq 0 \\
          -y_3 \leq 0 \\
          y_3 - 1 \leq 0 \\
          -z_1 \leq 0 
      \end{array}
    \right.
    \hspace{5em}
    \mu_{S} \defas \left\{
      \begin{array}{l}
        x_1 = 0 \\
        y_1 = 0 \\
        y_2 = \frac{1}{2} \\
        y_3 = \frac{1}{2} \\
        z_1 = 0
      \end{array}
    \right.
  \end{displaymath}
By splitting-on-demand, the \laint-solver adds the branch-and-bound lemmas 
\begin{displaymath}
  L = \left\{
    \begin{array}{l}
    (y_2 - \lfloor \frac{1}{2} \rfloor \leq 0) \lor (-y_2 + \lfloor \frac{1}{2} \rfloor + 1\leq 0)\\
    (y_3 - \lfloor \frac{1}{2} \rfloor \leq 0) \lor (-y_3 + \lfloor \frac{1}{2} \rfloor + 1\leq 0)
    \end{array}
  \right. = 
\left\{
    \begin{array}{l}
    (y_2  \leq 0) \lor (-y_2 +  1\leq 0)\\
    (y_3  \leq 0) \lor (-y_3 +  1\leq 0),
    \end{array}
\right.
\end{displaymath}
which are passed back to the \dpll engine. Suppose 
\dpll first ``decides'' $(y_3\le 0)$ (plus possibly some literal in
the first clause) invoking the layered \laint-solver. The
inconsistency of the branch is detected directly by the \larat-solver,
which produces the \larat-proof and the corresponding \larat-lemma:

    \begin{small}
      \begin{displaymath}
        P_1 \defas \vcenter{\infer{
          2 \leq 0
        }{
          \infer{
            -5 y_3 + 2 \leq 0
          }{
            \infer{
              5 y_2 - 5 y_3 + 2 \leq 0
            }{
              \infer{
                y_1 + 5 y_2 - 5 y_3 + 2 \leq 0
              }{
                \deduce{y_1 + 5 y_2 - 5 y_3 - 2 x_1 +2 \leq 0}{} &
                \deduce{2 \cdot (x_1 \leq 0)}{}
              } & 
              \deduce{-y_1 \leq 0}{}
            } & 
            \deduce{5 \cdot(-y_2 \leq 0)}{}
          } & 
          \deduce{5 \cdot (y_3 \leq 0)}{}
        }}
      \end{displaymath}
      \begin{displaymath}
          C_1 \defas \neg (y_1 + 5 y_2 - 5 y_3 - 2 x_1 + 2 \leq 0) \lor
                     \neg (x_1 \leq 0) \lor
                     \neg (-y_1 \leq 0) \lor
                     \neg (-y_2 \leq 0) \lor
                     \neg (y_3 \leq 0). 
      \end{displaymath}
    \end{small}

\noindent
Then \dpll unit-propagates $\neg (y_3 \le 0),(-y_3 + 1 \le 0)$ and
decides $(y_2\le 
0)$. As before, the \larat-solver is sufficient to detect the
inconsistency of the assignment, producing: 

    \begin{small}
      \begin{displaymath}
        P_2 \defas  \vcenter{\infer{
          2 \leq 0
        }{
          \infer{
            -5 y_2 + 2 \leq 0
          }{
            \infer{
              -5 y_2 + 5 y_3 - 3 \leq 0
            }{
              \infer{
                -y_1 - 5 y_2 + 5 y_3 - 3 \leq 0
              }{
                \deduce{- y_1 - 5 y_2 + 5 y_3 + 4 z_1 - 3 \leq 0}{} &
                \deduce{4 \cdot (-z_1 \leq 0)}{}
              } & 
              \deduce{y_1 \leq 0}{}
            } & 
            \deduce{5 \cdot(-y_3 + 1\leq 0)}{}
          } & 
          \deduce{5 \cdot (y_2 \leq 0)}{}
        }}
      \end{displaymath}
      \begin{displaymath}
          C_2 \defas \neg (- y_1 - 5 y_2 + 5 y_3 + 4 z_1 - 3 \leq 0) \lor
                     \neg (-z_1 \leq 0) \lor
                     \neg (y_1 \leq 0) \lor
                     \neg (-y_3 + 1\leq 0) \lor
                     \neg (y_2 \leq 0). \\
      \end{displaymath}
    \end{small}

\noindent 
Consequently, also $\neg (y_2 \le 0),(-y_2+1\le 0)$ are
unit-propagated. Likewise, the next step produces:

    \begin{small}
      \begin{displaymath}
        P_3 \defas \vcenter{\infer{
          2 \leq 0
        }{
          \infer{
            5 y_2 - 3 \leq 0
          }{
            \infer{
              5 y_2 - 5 y_3 + 2 \leq 0
            }{
              \infer{
                y_1 + 5 y_2 - 5 y_3 + 2 \leq 0
              }{
                \deduce{y_1 + 5 y_2 - 5 y_3 - 2 x_1 + 2 \leq 0}{} &
                \deduce{2 \cdot (x_1 \leq 0)}{}
              } & 
              \deduce{-y_1 \leq 0}{}
            } & 
            \deduce{5 \cdot(y_3 - 1\leq 0)}{}
          } & 
          \deduce{5 \cdot (-y_2 +1 \leq 0)}{}
        }} 
      \end{displaymath}
      \begin{displaymath}
          C_3 \defas \neg (y_1 + 5 y_2 - 5 y_3 - 2 x_1 + 2 \leq 0) \lor
                     \neg (x_1 \leq 0) \lor
                     \neg (-y_1 \leq 0) \lor
                     \neg (y_3 -1 \leq 0) \lor
                     \neg (- y_2 + 1 \leq 0).
      \end{displaymath}
    \end{small}

\noindent 
Then no more assignment can be generated, so
that \dpll returns \unsatres, and can produce a resolution proof $P$. 

If $S$ is partitioned into $A,B$, since the lemmas involve only
one variable and thus cannot be AB-mixed, then 
 an interpolant can be computed from the Boolean resolution proof $P$
and the \larat-proofs $P_1,P_2,P_3$ 
in the standard way %
with Algorithm~\ref{alg:SMT_general_schema}.
\end{exa}
} %
\noindent Thanks to the observation above, in order to be able to produce interpolants
with splitting on-demand the only thing we need is to make sure that 
we do not generate lemmas containing $AB$-mixed terms.~%
This is always the case for ``normal'' branch-and-bound lemmas 
(since they involve only one variable),
but this is not true in general for ``extended'' branch-and-bound lemmas
generated from proofs of unsatisfiability 
using the ``cuts from proofs'' algorithm of \cite{cav09_lia}.
The following example shows one such case.

\begin{exa}
  Let $A$ and $B$ be defined as
  \begin{displaymath}
    \begin{split}
      A \defas (y - 2x \leq 0) \land (2x - y \leq 0), \ \ \
      B \defas (y - 2z - 1 \leq 0) \land (2z + 1 - y \leq 0)
    \end{split}
  \end{displaymath}
  When solving $A \land B$ using extended branch and bound, we might 
  generate the following $AB$-mixed lemma:
  \begin{math}
    (x - z \leq 0) \lor (-x +z + 1 \leq 0).
  \end{math}
\end{exa}

\noindent Since we want to be able to reuse the Boolean interpolation algorithm 
also for splitting on-demand, 
we want to avoid generating $AB$-mixed lemmas.
However, we would still like to exploit the cuts from proofs algorithm of \cite{cav09_lia} as much as possible.
We describe how we do this in the following.

\subsubsection{Interpolation with the cuts from proofs algorithm}
The core of the cuts from proofs algorithm is the identification of the 
\emph{defining constraints} of the current solution 
of the rational relaxation of the input set of \laint constraints.
A defining constraint is an input constraint $\sum_i c_i v_i + c \bowtie 0$ 
(where $\bowtie\,\in \{ \leq, = \}$) such that $\sum_i c_i v_i + c$ 
evaluates to zero under the current solution for the rational relaxation of the problem.
After having identified the defining constraints $D$, the cuts from proofs algorithm
checks the satisfiability of the system of Diophantine equations
$D_E \defas \{ \sum_i c_i v_i + c = 0 ~|~ (\sum_i c_i v_i + c \bowtie 0) \in D \}$.
If $D_E$ is unsatisfiable, 
then it is possible to generate a proof of unsatisfiability for it.
The root of such proof is an equation
$\sum_i c'_i v_i + c' = 0$
such that the GCD $g$ of the $c'_i$'s does not divide $c'$.
From such equation, 
it is generated
the extended branch and bound lemma:
{%
\begin{displaymath}
  (\sum_i \dfrac{c'_i}{g} v_i \leq \left\lceil \dfrac{-c'}{g} \right\rceil -1) \lor
  (\left\lceil \dfrac{-c'}{g} \right\rceil \leq \sum_i \dfrac{c'_i}{g} v_i).
\end{displaymath}
\begin{exa}\label{ex:cut_from_proof}%
  Consider the following set of \laint-constraints and its rational relaxation solution $\mu_{S}$
  \begin{displaymath}
    S \defas \left\{
      \begin{array}{l}
        5v_1 - 5v_2 - v_3 -3 \leq 0 \\
        -5v_1 + 5v_2 + v_3 + 2 \leq 0 \\
        v_3 \leq 0 \\
        -v_3 \leq 0
      \end{array}
    \right.
        \hspace{5em}
        \mu_{S} \defas \left\{
          \begin{array}{l}
            v_1 = 0 \\
            v_2 = -\frac{2}{5} \\
            v_3 = 0
          \end{array}
        \right.
  \end{displaymath}
  The set of defining constraints $D$ for $\langle S,\mu_S\rangle$ is then:
  \begin{displaymath}
    D \defas \left\{
      \begin{array}{l}
        -5v_1 + 5v_2 + v_3 + 2 \leq 0 \\
        v_3 \leq 0 \\
        -v_3 \leq 0,
      \end{array}
    \right.
  \end{displaymath}
  resulting in the following inconsistent system of Diophantine equations $D_E$:
  \begin{displaymath}
    D_E \defas \left\{
      \begin{array}{l}
        -5v_1 + 5v_2 + v_3 + 2 = 0 \\
        v_3 = 0
      \end{array}
    \right.
  \end{displaymath}
  The Diophantine equations handler generates $-5v_1 + 5v_2 + 2 = 0$ as proof of unsatisfiability for $D_E$,
  resulting in the following branch-and-bound lemma:
  \begin{equation}\label{eq:extended_bb_lemma}
    (-v_1 + v_2 \leq \left\lceil \dfrac{-2}{5}  \right\rceil -1) \lor (\left\lceil\dfrac{-2}{5}\right\rceil\leq -v_1 + v_2) \text{, or equivalently } (-v_1+v_2+1\leq 0) \lor (v_1-v_2\leq 0)
  \end{equation}
  After adding \eqref{eq:extended_bb_lemma} to \dpll, 
  the \larat-solver detects the \larat-inconsistency of both
  $S \cup (-v_1 + v_2 + 1 \leq 0)$ and $S \cup (v_1 - v_2 \leq 0)$.
\end{exa}
}
If $\sum_i \dfrac{c'_i}{g} v_i$ is not $AB$-mixed, we can generate the above lemma also when computing interpolants.
If $\sum_i \dfrac{c'_i}{g} v_i$ is $AB$-mixed, instead, we generate a different lemma,
still exploiting the unsatisfiability of (the equations corresponding to) the defining constraints.
Since $D_E$ is unsatisfiable, we know that the current rational solution $\mu$
is not compatible with the current set of defining constraints.
If the defining constraints were all equations,
the submodule for handling Diophantine equations would have detected the conflict.
Therefore, there is at least one defining constraint 
$\sum_i \bar{c}_i v_i + \bar{c} \leq 0$.
Our idea is that of \emph{splitting} this constraint into 
$(\sum_i \bar{c}_i v_i + \bar{c}+1 \leq 0)$ and $(\sum_i \bar{c}_i v_i + \bar{c} = 0)$, by generating the lemma
\begin{displaymath}
\neg(\sum_i \bar{c}_i v_i + \bar{c} \leq 0) \lor 
(\sum_i \bar{c}_i v_i + \bar{c}+1 \leq 0) \lor 
(\sum_i \bar{c}_i v_i + \bar{c} = 0).
\end{displaymath}
In this way, we are either ``moving away'' from the current bad rational solution $\mu$ 
(when $(\sum_i \bar{c}_i v_i + \bar{c}+1 \leq 0)$ is set to true),
or we are forcing one more element of the set of defining constraints to be an equation 
(when $(\sum_i \bar{c}_i v_i + \bar{c} = 0)$ is set to true):
if we repeat the splitting, then, eventually all the defining constraints for the bad solution $\mu$ 
will be equations, thus allowing the Diophantine equations handler to detect the conflict
without the need of generating more branch-and-bound lemmas.
Since the set of defining constraints is a subset of the input constraints, 
lemmas generated in this way will never be $AB$-mixed.

It should be mentioned that this procedure is very similar to the algorithm used in the recent work \cite{lpar10_interpolation}
for avoiding the generation of $AB$-mixed cuts.
However, the criterion used to select which inequality to split and how to split it is different
(in \cite{lpar10_interpolation} such inequality is selected among those that
are violated by the closest integer solution to the current rational solution).
Moreover, we don't do this systematically, 
but rather only if the cuts from proofs algorithm is not able to generate 
a non-$AB$-mixed lemma by itself.
In a sense, the approach of \cite{lpar10_interpolation} is ``pessimistic'' 
in that it systematically excludes certain kinds of cuts, 
whereas our approach is more ``optimistic''.

\subsubsection{Interpolation for the internal branch-and-bound module}

From the point of view of interpolation
the subdivision of the branch-and-bound module in an ``internal'' and an ``external'' part
poses no difficulty.
The only difference between the two is that in the former the case splits 
are performed by the \laint-solver instead of \dpll.
However, we can still treat such case splits as if they were performed by \dpll,
build a Boolean resolution proof for the \laint-conflicts discovered by the internal branch-and-bound procedure,
and then apply the propositional interpolation algorithm as in the case of splitting on-demand.

{%
More specifically, a \emph{branch-and-bound proof} 
is a tree in which the leaves are \larat-proofs of unsatisfiability,
the root is a \laint-conflict set,
and each internal node has two children that are labeled with two ``complementary'' atoms $(v - n \leq 0)$ and $(-v + n + 1 \leq 0)$.
From a branch-and-bound proof, a resolution proof for the \laint-lemma corresponding to the root \laint-conflict set can be generated 
by replacing each leaf \larat-proof $P$ with the corresponding \larat-lemma $C$,
and by introducing, for each internal node, 
a branch-and-bound lemma $(v-n\leq 0) \lor (-v+n+1 \leq 0)$ and two resolution steps,
according to the following pattern:

\begin{small}
  \begin{displaymath}
    \infer[\text{pivot on~}(-v+n+1\leq 0)]{P}{
      \infer[\text{pivot on~}(v-n\leq 0)]{\cdot}
      {(v-n\leq 0) \lor (-v+n+1\leq 0) & P_l}
      & P_r}
  \end{displaymath}
\end{small}

The following example shows how this is done.
} %
{%
  \begin{exa}
Consider the same set of $S$ as in Example~\ref{ex:splitting-on-demand},
partitioned as follows: 
    \begin{displaymath}
      A \defas \left\{
        \begin{array}{l}
          (y_1 + 5 y_2 - 5 y_3 - 2 x_1 + 2 \leq 0) \\
          (x_1 \leq 0) \\
          (y_1 \leq 0) \\
          (y_2 - 2 \leq 0) \\
          (y_3 - 1 \leq 0)
        \end{array}
      \right.
      \hfill
      B \defas \left\{
        \begin{array}{l}
          (- y_1 - 5 y_2 + 5 y_3 + 4 z_1 - 3 \leq 0) \\
          (-z_1 \leq 0) \\
          (-y_1 \leq 0) \\
          (-y_2 \leq 0) \\
          (-y_3 \leq 0)
        \end{array}
      \right.
    \end{displaymath}

    A branch-and-bound proof $P$ that shows the unsatisfiability of $A
    \land B$ is the following:~\footnote{%
The \larat-proofs $P_i$ and \larat-lemmas $C_i$ are the same as in
Example~\ref{ex:splitting-on-demand}; they are reported here for convenience.}
    \begin{small}
    \begin{displaymath}
      P \defas \vcenter{\infer[\langle (y_3 \leq 0), (-y_3+1 \leq 0) \rangle]{
        \bot
      }{
        \infer[\langle (y_2 \leq 0), (-y_2+1\leq 0) \rangle]{
          \cdot
        }{\deduce{P_2}{} & \deduce{P_3}{}}
        &
        \deduce{P_1}{}
      }}
    \end{displaymath}
      \begin{displaymath}
        P_1 \defas \vcenter{\infer{
          2 \leq 0
        }{
          \infer{
            -5 y_3 + 2 \leq 0
          }{
            \infer{
              5 y_2 - 5 y_3 + 2 \leq 0
            }{
              \infer{
                y_1 + 5 y_2 - 5 y_3 + 2 \leq 0
              }{
                \deduce{y_1 + 5 y_2 - 5 y_3 - 2 x_1 +2 \leq 0}{} &
                \deduce{2 \cdot (x_1 \leq 0)}{}
              } & 
              \deduce{-y_1 \leq 0}{}
            } & 
            \deduce{5 \cdot(-y_2 \leq 0)}{}
          } & 
          \deduce{5 \cdot (y_3 \leq 0)}{}
        }}
      \end{displaymath}
    \end{small}
    \begin{small}
      \begin{displaymath}
        P_2 \defas \vcenter{\infer{
          2 \leq 0
        }{
          \infer{
            -5 y_2 + 2 \leq 0
          }{
            \infer{
              -5 y_2 + 5 y_3 - 3 \leq 0
            }{
              \infer{
                -y_1 - 5 y_2 + 5 y_3 - 3 \leq 0
              }{
                \deduce{- y_1 - 5 y_2 + 5 y_3 + 4 z_1 - 3 \leq 0}{} &
                \deduce{4 \cdot (-z_1 \leq 0)}{}
              } & 
              \deduce{y_1 \leq 0}{}
            } & 
            \deduce{5 \cdot(-y_3 + 1\leq 0)}{}
          } & 
          \deduce{5 \cdot (y_2 \leq 0)}{}
        }}
      \end{displaymath}
    \end{small}
    \begin{small}
      \begin{displaymath}
        P_3 \defas \vcenter{\infer{
          2 \leq 0
        }{
          \infer{
            5 y_2 - 3 \leq 0
          }{
            \infer{
              5 y_2 - 5 y_3 + 2 \leq 0
            }{
              \infer{
                y_1 + 5 y_2 - 5 y_3 + 2 \leq 0
              }{
                \deduce{y_1 + 5 y_2 - 5 y_3 - 2 x_1 + 2 \leq 0}{} &
                \deduce{2 \cdot (x_1 \leq 0)}{}
              } & 
              \deduce{-y_1 \leq 0}{}
            } & 
            \deduce{5 \cdot(y_3 - 1\leq 0)}{}
          } & 
          \deduce{5 \cdot (-y_2 +1 \leq 0)}{}
        }}
      \end{displaymath}
    \end{small}

    \medskip
    A corresponding resolution proof, 
    is then:

    \begin{small}
      \begin{displaymath}
        \begin{array}{l}
          C_1 \defas \neg (y_1 + 5 y_2 - 5 y_3 - 2 x_1 + 2 \leq 0) \lor
                     \neg (x_1 \leq 0) \lor
                     \neg (-y_1 \leq 0) \lor
                     \neg (-y_2 \leq 0) \lor
                     \neg (y_3 \leq 0) \\
          C_2 \defas \neg (- y_1 - 5 y_2 + 5 y_3 + 4 z_1 - 3 \leq 0) \lor
                     \neg (-z_1 \leq 0) \lor
                     \neg (y_1 \leq 0) \lor
                     \neg (-y_3 + 1\leq 0) \lor
                     \neg (y_2 \leq 0) \\
          C_3 \defas \neg (y_1 + 5 y_2 - 5 y_3 - 2 x_1 + 2 \leq 0) \lor
                     \neg (x_1 \leq 0) \lor
                     \neg (-y_1 \leq 0) \lor
                     \neg (y_3 -1 \leq 0) \lor
                     \neg (- y_2 + 1 \leq 0)
        \end{array}
      \end{displaymath}
    \end{small}

    \begin{footnotesize}
      \begin{displaymath}
        \infer{
              \efrac{
                \neg (- y_1 - 5 y_2 + 5 y_3 + 4 z_1 - 3 \leq 0) \lor
                \neg (-z_1 \leq 0) \lor
                \neg (y_1 \leq 0) \lor
                }{
                \neg (y_1 + 5 y_2 - 5 y_3 - 2 x_1 + 2 \leq 0) \lor
                \neg (x_1 \leq 0) \lor
                \neg (-y_1 \leq 0) \lor
                \neg (y_3 -1 \leq 0) \lor
                \neg (-y_2 \leq 0)
              }
        }{
          \infer{
              \efrac{
                \neg (- y_1 - 5 y_2 + 5 y_3 + 4 z_1 - 3 \leq 0) \lor
                \neg (-z_1 \leq 0) \lor
                \neg (y_1 \leq 0) \lor
                }{
                \neg (y_1 + 5 y_2 - 5 y_3 - 2 x_1 + 2 \leq 0) \lor
                \neg (x_1 \leq 0) \lor
                \neg (-y_1 \leq 0) \lor
                \neg (y_3 -1 \leq 0) \lor
                (y_3 \leq 0)
              }
          }{
            (y_3 \leq 0) \lor (-y_3 + 1 \leq 0) &
            \infer{
              \efrac{
                \neg (- y_1 - 5 y_2 + 5 y_3 + 4 z_1 - 3 \leq 0) \lor
                \neg (-z_1 \leq 0) \lor
                \neg (y_1 \leq 0) \lor
                \neg (-y_3 + 1\leq 0) \lor
                }{
                \neg (y_1 + 5 y_2 - 5 y_3 - 2 x_1 + 2 \leq 0) \lor
                \neg (x_1 \leq 0) \lor
                \neg (-y_1 \leq 0) \lor
                \neg (y_3 -1 \leq 0)
              }
              }{
              \infer{
                \efrac{
                \neg (- y_1 - 5 y_2 + 5 y_3 + 4 z_1 - 3 \leq 0) \lor
                \neg (-z_1 \leq 0) \lor
                }{
                \neg (y_1 \leq 0) \lor
                \neg (-y_3 + 1\leq 0) \lor
                (-y_2 + 1 \leq 0)
              }
              }{
                (y_2 \leq 0) \lor (-y_2 + 1 \leq 0) &
                C_2
              } & C_3
            }
          } & C_1
        }
      \end{displaymath}          
    \end{footnotesize}

    \noindent 
    where $C_1$, $C_2$ and $C_3$ are the \larat-lemmas corresponding to the \larat-proofs $P_1$, $P_2$ and $P_3$ respectively.
    Applying Algorithm~\ref{alg:SMT_general_schema} to this proof,
    and considering all the branch-and-bound atoms as part of $B$ (since they are all on $AB$-common variables),
    results in the following \laint-interpolant $I$ for the \laint-lemma corresponding to the root of the proof:
    \begin{displaymath}
      I \defas (y_1 \leq 0) \land 
      (y_1 + 5 y_2 -5 y_3 + 2 \leq 0) \land
      (y_1 + 5 y_2 - 3 \leq 0).
    \end{displaymath}
  \end{exa}
} %

\section{A novel general interpolation technique for inequalities}
\label{sec:itp_ceiling}

The use of the Strengthen rule allows us to produce interpolants 
with very little modifications to the \laint-solver
(we only need to enable the generation of cutting-plane proofs),
which in turn result in very little overhead \emph{at search time}.
However, the Strengthen rule might cause a very significant overhead 
\emph{when generating the interpolant} from a proof of unsatisfiability.
In fact, even a single Strengthen application results in a disjunction 
whose size is proportional to the \emph{value of the constant} $k$ in the rule.
The following example, taken from \cite{lpar10_interpolation}, illustrates the problem.

\begin{exa}\label{ex:strengthen_blowup}
  Consider the following (parametric) interpolation problem
  \cite{lpar10_interpolation}: 
  \begin{displaymath}
  \begin{split}
    A \defas (-y_1 -2nx_1 -n+1 \leq 0) \land (y_1 + 2nx_1 \leq 0) \\
    B \defas (-y_1 -2nz_1 + 1 \leq 0) \land (y_1 + 2nz_1 - n \leq 0)
  \end{split}
  \end{displaymath}
  where the parameter $n$ is an integer constant greater than 1.  
Using the rules of \sref{sec:interpolation_congr},
we can construct the following annotated cutting-plane proof of unsatisfiability:
\begin{small}
\begin{displaymath}\textstyle
    \infer{
      1 \leq 0 ~
      \begin{array}{l}
      [\{\itpelem{j-n+1 \leq 0}{\exists x_2.(y_1 + 2nx_2 + j = 0)}~|~0 \leq j < {2n-1}\} \cup \\
      \phantom{[}\{\itpelem{({2n - 1})-n+1 \leq 0}{\top}\}]
       \end{array}
    }{
      \infer{
	\deduce{
          \begin{array}{l}
            [\{\itpelem{y_1 + 2nx_1 + j \leq 0}{\\~~~~\exists x_2.(y_1 + 2nx_2 + j = 0)}~|~ \\
            ~~~~0 \leq j < {2n-1}\} \cup \\
            \{\itpelem{y_1 + 2nx_1 + 2n - 1 \leq 0}{\top}\}]
          \end{array}
        }{2nx_1 - 2nz_1 + 1 + (2n-1) \leq 0}
      }{
	\infer{
	  \deduce{[\{\itpelem{y_1 + 2nx_1 \leq 0}{\top}\}]}{2nx_1 - 2nz_1 + 1 \leq 0}
	}{
	  \deduce{[\{\itpelem{y_1 + 2nx_1 \leq 0}{\top}\}]}{y_1 + 2nx_1 \leq 0} &
	  \deduce{[\{\itpelem{0 \leq 0}{\top}\}]}{-y_1 - 2nz_1 + 1 \leq 0}
	}
      } &
      \infer{
	\deduce{[\{\itpelem{-y_1 - 2nx_1 - n + 1 \leq 0}{\top}\}]}{-2nx_1 + 2nz_1 -2n + 1 \leq 0}
      }{
	\deduce{[\{\itpelem{-y_1 - 2nx_1 - n + 1 \leq 0}{\top}\}]}{-y_1 - 2nx_1 - n + 1 \leq 0} &
	\deduce{[\{\itpelem{0 \leq 0}{\top}\}]}{y_1 + 2nz_1 - n \leq 0}
      }
    }  
\end{displaymath}
\end{small}
By observing that $(j-n+1 \leq 0) \models \bot$ when $j \geq n$,
the generated interpolant is:
\vspace{-1.5ex}
\begin{displaymath}
    (y_1 =_{2n} -n + 1) \lor (y_1 =_{2n} -n + 2) \lor \ldots \lor (y_1 =_{2n} 0),
\end{displaymath}
whose size is linear in $n$, and thus exponential wrt. the size of the input problem.
In fact,
in \cite{lpar10_interpolation}, it is said that this is the only (up to equivalence)
interpolant for $(A, B)$ that can be obtained by using only interpreted symbols in the signature $\Sigma \defas \{ =, \leq, +, \cdot \} \cup \mathbb{Z} \cup \{ =_g | g \in \mathbb{Z}^{>0} \}$.
\end{exa}

\noindent In order to overcome this drawback,
we present a novel and very effective way of computing interpolants in
\laint, which is inspired by a result by Pudl\'ak
\cite{pudlak_interpolation}. The key idea is \emph{to extend both the
  signature and the domain} of the theory by explicitly introducing the
\emph{ceiling function} $\lceil \cdot \rceil$ and by allowing
non-variable terms to be non-integers.

As in Section \sref{sec:interpolation}, we use the annotated rules {Hyp-A},
{Hyp-B} and {Comb}. 
However, in this case the annotations are {\em single} inequalities in the
form $(t\le 0)$
rather than (possibly large) sets of inequalities and equalities.
Moreover, we replace the Strengthen rule with the equivalent Division rule:
\begin{description}
\item[Division] 
  \begin{displaymath}
\textstyle
\hspace{-.5cm}
    \infer{
      (A, B) \vdash 
             \sum_i \dfrac{a_i}{d} x_i + \sum_j \dfrac{c_j}{d} y_j + \sum_k \dfrac{b_k}{d} z_k + \left\lceil \dfrac{c}{d} \right\rceil \leq 0
           \  [\sum_i \dfrac{a_i}{d} x_i + \left\lceil \dfrac{\sum_j c'_j y_j + c'}{d} \right\rceil \leq 0]
    }{
      (A, B) \vdash 
             \sum_i a_i x_i + \sum_j c_j y_j + \sum_k b_k z_k + c \leq 0
           \  [\sum_i a_i x_i + \sum_j c'_j y_j + c' \leq 0]
    }
  \end{displaymath}
\noindent
  where:
  \begin{iteMize}{$-$}
  \item $x_i \not\in B$, $y_j \in A \cap B$, $z_k \not\in A$
  \item $d > 0$ divides all the $a_i$'s, $c_j$'s and $b_k$'s 
  \end{iteMize}
\end{description}

\noindent
As before, if we ignore the presence of annotations,
 the rules {Hyp-A}, {Hyp-B}, {Comb} and
{Division} form a complete proof systems for \laint \cite{ilp_book}.
Notice also that all the rules {Hyp-A}, {Hyp-B}, {Comb} and
{Division} preserve the following invariant:  the coefficients $a_i$ of 
the A-local variables are always the same for the implied inequality and its
 annotation. 
This makes the Division rule always applicable.
Therefore, the above rules can be used to annotate any cutting-plane proof.
{In particular, this means that our new technique 
can be applied also to proofs generated by other \laint techniques 
used in modern \smt solvers, such as
those based on Gomory cuts
or on the Omega test~\cite{omega}.}

\begin{defi}
\label{def:valid_sequent_ceiling}
  An annotated sequent $(A, B) \vdash (t \leq 0) [(t' \leq 0)]$ 
  is \emph{valid} when:
  \begin{enumerate}[(1)]
  \item $A \models (t' \leq 0)$;
  \item $B \models (t - t' \leq 0)$;
  \item $t' \preceq A$ and $(t - t') \preceq B$.
  \end{enumerate}
\end{defi}

\begin{thm}
\label{teo:ceiling_int}
  All the interpolating rules preserve the validity of the sequents.
\end{thm}
{%
\proof 
  The theorem can be easily proved for Hyp-A, Hyp-B and Comb. 
  Therefore, here we focus only on Division.\hfill
  \begin{enumerate}[(1)]
  \item By hypothesis, $A \models \sum_i a_i x_i + \sum_j c'_j y_j + c' \leq 0$.
    Since $d > 0$, we have that 
    \begin{displaymath}
      A \models \dfrac{\sum_i a_i x_i + \sum_j c'_j y_j + c'}{d} \leq 0.
    \end{displaymath}
    From the definition of ceiling, therefore
    \begin{displaymath}
      A \models \left\lceil \dfrac{\sum_i a_i x_i + \sum_j c'_j y_j + c'}{d} \right\rceil \leq 0.
    \end{displaymath}
    Since $d$ divides the $a_i$'s by hypothesis, 
    $\dfrac{\sum_i a_i x_i}{d}$ is an integer, and since 
    $\lceil n + x \rceil \equiv n + \lceil x \rceil$ if $n$ is an integer,
    we have that
    \begin{displaymath}
      A \models \sum_i \dfrac{a_i}{d} x_i + \left\lceil \dfrac{\sum_j c'_j y_j + c'}{d} \right\rceil \leq 0.
    \end{displaymath}
  \item By hypothesis, $B \models (\sum_i a_i x_i + \sum_j c_j y_j +
    \sum_k b_k z_k + c) - (\sum_i a_i x_i + \sum_j c'_j y_j + c') \leq 0$.
    Since $d > 0$, then
    \begin{displaymath}
      B \models \dfrac{(\sum_i a_i x_i + \sum_j c_j y_j + \sum_k b_k z_k + c)}{d} - \dfrac{(\sum_i a_i x_i + \sum_j c'_j y_j + c')}{d} \leq 0
    \end{displaymath}
    and thus
    \begin{displaymath}
      B \models \left\lceil \dfrac{(\sum_i a_i x_i + \sum_j c_j y_j + \sum_k b_k z_k + c)}{d} - \dfrac{(\sum_i a_i x_i + \sum_j c'_j y_j + c')}{d} \right\rceil \leq 0.
    \end{displaymath}
    By observing that $\lceil x - y \rceil \geq \lceil x \rceil -
    \lceil y \rceil$, we have: 
    \begin{displaymath}
      B \models \left\lceil \dfrac{(\sum_i a_i x_i + \sum_j c_j y_j + \sum_k b_k z_k + c)}{d} \right\rceil -  \left\lceil \dfrac{(\sum_i a_i x_i + \sum_j c'_j y_j + c')}{d} \right\rceil \leq 0.
    \end{displaymath}
    By observing  that $\lceil n + x \rceil \equiv n + \lceil x
    \rceil$ when $n$ is an integer, we have finally:
    \begin{displaymath}
      B \models (\sum_i \dfrac{a_i}{d} x_i + \sum_j \dfrac{c_j}{d} y_j + \sum_k \dfrac{b_k}{d} z_k + \left\lceil \dfrac{c}{d} \right\rceil) - (\sum_i \dfrac{a_i}{d} x_i + \left\lceil \dfrac{\sum_j c'_j y_j + c'}{d} \right\rceil) \leq 0.
    \end{displaymath}
  \item Follows directly from the hypothesis.
    \qed
  \end{enumerate}
  }
\begin{cor}
\label{cor:ceiling_int}
  If we can derive a valid sequent $(A, B) \vdash c \leq 0 [t \leq 0]$ with 
  $c > 0$, then $(t \leq 0)$ is an interpolant for $(A, B)$.
\end{cor}
{%
\proof\hfill
  \begin{enumerate}[(1)]
  \item $\mathbf{A \models (t \leq 0).}$ Trivial from
    Definition~\ref{def:valid_sequent_ceiling} and
    Theorem~\ref{teo:ceiling_int}. 
  \item $\mathbf{B \wedge (t  \leq 0) \models \bot.}$
From Definition~\ref{def:valid_sequent_ceiling} and
Theorem~\ref{teo:ceiling_int} we have $B\models (c-t\le 0)$.
Since $c>0$, then  $B\models (-t< 0) \equiv (t> 0)$, so that $B \wedge
(t\le 0)\models \bot$.
  \item $\mathbf{(t\leq 0) \preceq A}$ and $\mathbf{(t\leq 0) \preceq B.}$ Trivial from
    Definition~\ref{def:valid_sequent_ceiling} and
    Theorem~\ref{teo:ceiling_int}.  \qed
  \end{enumerate}
  }
\begin{exa}
  Consider the following interpolation problem:
  \begin{displaymath}
  \begin{split}
    A \defas (y_1 = 2x_1), \quad\quad  
    & B \defas (y_1 = 2z_1 + 1).
  \end{split}
  \end{displaymath}
  The following is an annotated cutting-plane proof of unsatisfiability for $A \land B$:
  \begin{small}
  \begin{displaymath}
    \infer{
      1 \leq 0 [-y_1 + 2 \left\lceil \frac{y_1}{2}\right\rceil \leq 0]
    }{
      \infer{
	z_1 - x_1 + 1 \leq 0 [-x_1 + \left\lceil \frac{y_1}{2}\right\rceil \leq 0]
      }{
	\infer{
	  2z_1 - 2x_1 + 1 \leq 0 [y_1 - 2x_1 \leq 0]
	}{
	  \infer{y_1 - 2x_1 \leq 0 [y_1 - 2x_1 \leq 0]}{y_1 = 2x_1} &
	  \infer{2z_1 + 1 - y_1 \leq 0 [0 \leq 0]}{y_1 = 2z_1+1}
	}
      } &
      \infer{
	2x_1 - 2z_1 - 1 \leq 0 [2x_1 - y_1 \leq 0]
      }{
	\infer{\deduce{[2x_1 - y_1 \leq 0]}{2x_1 - y_1 \leq 0}}{y_1 = 2x_1} &
	\infer{\deduce{[0 \leq 0]}{y_1 - 2z_1 - 1 \leq 0}}{y_1 = 2z_1 + 1}
      }
    }
  \end{displaymath}
  \end{small}

\noindent
  Then, $(-y_1 + 2 \left\lceil \dfrac{y_1}{2}\right\rceil \leq 0)$
  is an interpolant for $(A, B)$.
\end{exa}
\noindent Using the ceiling function,
we do not incur in any blowup of the size of the generated interpolant
wrt. the size of the proof of unsatisfiability.%
\footnote{However, we remark that, in general, cutting-plane proofs of unsatisfiability can be exponentially large wrt. the size of the input problem \cite{ilp_book,pudlak_interpolation}.}
In particular, by using the ceiling function we might produce interpolants
which are up to exponentially smaller than those generated using modular equations.
{The intuition is that the use of the ceiling function in the annotation
of the Division rule allows for expressing \emph{symbolically} 
the case distinction that the Strengthen rule of \sref{sec:interpolation_congr}
was expressing \emph{explicitly} as a disjunction of modular equations.}

\begin{exa}
  Consider again the parametric interpolation problem of Example~\ref{ex:strengthen_blowup}:
  \vspace{-1.5ex}
  \begin{displaymath}
  \begin{split}
    A \defas (-y_1 -2nx_1 -n+1 \leq 0) \land (y_1 + 2nx_1 \leq 0) \\
    B \defas (-y_1 -2nz_1 + 1 \leq 0) \land (y_1 + 2nz_1 - n \leq 0)
  \end{split}
  \end{displaymath}
  Using the ceiling function, we can generate the following annotated proof:
  \begin{small}
  \begin{displaymath}\textstyle
    \infer{
      1 \leq 0 ~[2n \left\lceil \frac{y_1}{2n}\right\rceil -y_1 -n + 1 \leq 0]
    }{
      \infer{
        \deduce{\phantom{2n\cdot(}[x_1 + \left\lceil \frac{y_1}{2n}\right\rceil \leq 0]}{x_1 - z_1 + 1 \leq 0}
      }{
	\infer{
	  \deduce{[y_1 + 2nx_1 \leq 0]}{2nx_1 - 2nz_1 + 1 \leq 0}
	}{
	  \deduce{[y_1 + 2nx_1 \leq 0]}{y_1 + 2nx_1 \leq 0} &
	  \deduce{[0 \leq 0]}{-y_1 - 2nz_1 + 1 \leq 0}
	}
      } &
      \infer{
	\deduce{[ -y_1 - 2nx_1 - n + 1 \leq 0 ]}{-2nx_1 + 2nz_1 -2n + 1 \leq 0}
      }{
	\deduce{[ -y_1 - 2nx_1 - n + 1 \leq 0 ]}{-y_1 - 2nx_1 - n + 1 \leq 0} &
	\deduce{[0 \leq 0]}{y_1 + 2nz_1 - n \leq 0}
      }
    }
  \end{displaymath}
  \end{small}
  The interpolant corresponding to such proof is then
  $(2n \left\lceil \dfrac{y_1}{2n}\right\rceil -y_1 -n + 1 \leq 0)$,
  whose size is linear in the size of the input.
\end{exa}

\subsection{Solving and interpolating formulas with ceilings}

Any \smt solver supporting \laint can be easily extended to support formulas containing ceilings.
In fact, we notice that we can
eliminate ceiling functions from a formula $\varphi$
with a simple preprocessing step as follows:

\begin{enumerate}[(1)]
\item Replace every term $\lceil t_i \rceil$ occurring in $\varphi$ with a fresh integer variable $x_{\lceil t_i \rceil}$;
\item Set $\varphi$ to $\varphi \land \bigwedge_i \{ (x_{\lceil t_i \rceil} - 1 < t_i \leq x_{\lceil t_i \rceil}) \}$.
\end{enumerate}

Moreover, we remark that
for using ceilings we must only be able to \emph{represent} non-variable terms with rational coefficients,
but we don't need to extend our \laint-solver to support Mixed Rational/Integer Linear Arithmetic.
This is because, after the elimination of ceilings performed during preprocessing,
we can multiply both sides of the introduced constraints 
$(x_{\lceil t_i \rceil} -1 < t_i)$ and $(t_i \leq x_{\lceil t_i \rceil})$ 
by the least common multiple of the rational coefficients in $t_i$,
thus obtaining two \laint-inequalities.

For interpolation, it is enough to preprocess $A$ and $B$ separately,
so that the elimination of ceilings will not introduce variables common to $A$ and $B$.

\subsection{Generating sequences of interpolants}

{
One of the most important applications of interpolation in Formal Verification
is abstraction refinement \cite{AbstractionsFromProofs,McMillanCAV06}. 
In such setting, every input problem $\phi$ has the form 
$\phi \defas \phi_1 \wedge \ldots \wedge \phi_n$, 
and the interpolating solver is asked to compute a \emph{sequence} of interpolants 
$I_1,\ldots,I_{n-1}$ 
corresponding to different partitions of $\phi$ into $A_i$ and $B_i$, 
such that $\forall i,~~ A_i \defas \phi_1 \wedge \ldots \wedge \phi_i, \text{~~and~~} B_i \defas \phi_{i+1} \wedge \ldots \wedge \phi_n$.
Moreover, $I_1,\ldots,I_{n-1}$ should be related by the following:
\begin{equation}
  \label{eq:itp:multiple_interpolants}
  I_i \wedge \phi_{i+1} \models I_{i+1}
\end{equation}
}

{
As stated (without proof) in \cite{AbstractionsFromProofs}, 
a sufficient condition for \eqref{eq:itp:multiple_interpolants} to hold 
is that all the $I_i$'s are computed from the same proof of unsatisfiability
for $\phi$.
In our previous work \cite{tocl_interpolation} (Theorem~6.6, page~7:46),
we have formally proved that such sufficient condition is valid 
for \emph{every} $\smtt$-proof of unsatisfiability,
\emph{independently} of the background theory \T.
By observing that all the techniques that we have described in this article
do not involve modifications/manipulations of the proofs of unsatisfiability,
we can immediately conclude that 
this approach can be applied without modifications also in our context,
for computing sequences of interpolants for \laint-formulas
using our interpolation algorithms.
}

\section{Experimental evaluation}
\label{sec:expeval}

The techniques presented in previous sections have been implemented
within the \mathsat 5 SMT solver \cite{mathsat5}.
In this section, we experimentally evaluate our approach.

\subsection{Experiments on large SMT formulas}
\label{sec:benchmarks}

{%
In the first part of our experimental analysis,
we evaluate the performance of our techniques 
on relatively-large formulas taken from the set of benchmark instances
in the QF\_LIA (``quantifier-free \laint'') category of the SMT-LIB.}%
\footnote{\url{http://smtlib.org}}
More specifically, we have selected the subset of \laint-unsatisfiable instances
whose rational relaxation is (easily) satisfiable,
so that \laint-specific interpolation techniques are put under stress.
In order to generate interpolation problems, 
we have split each of the collected instances in two parts $A$ and $B$,
by collecting about 40\%
and making sure that $A$ contains some symbols not occurring in $B$ 
(so that $A$ is never a ``trivial'' interpolant).
In total, our benchmark set consists of 513 instances.%

We have run the experiments on a machine with 
a 2.6 GHz Intel Xeon processor, 
16 GB of RAM and 6 MB of cache,
running Debian GNU/Linux 5.0.
We have used a time limit of 1200 seconds and a memory limit of 3 GB.

\begin{figure}[t]
  \begin{tabular}{cccc}
    & \iprincess & \hspace{1em}\opensmtitp & \hspace{2em}\smtinterpol \\
    \rotatebox{90}{\hspace{0.5em}\mathsatcongr} &
    \includegraphics[scale=0.45, bb=126 63 360 285]{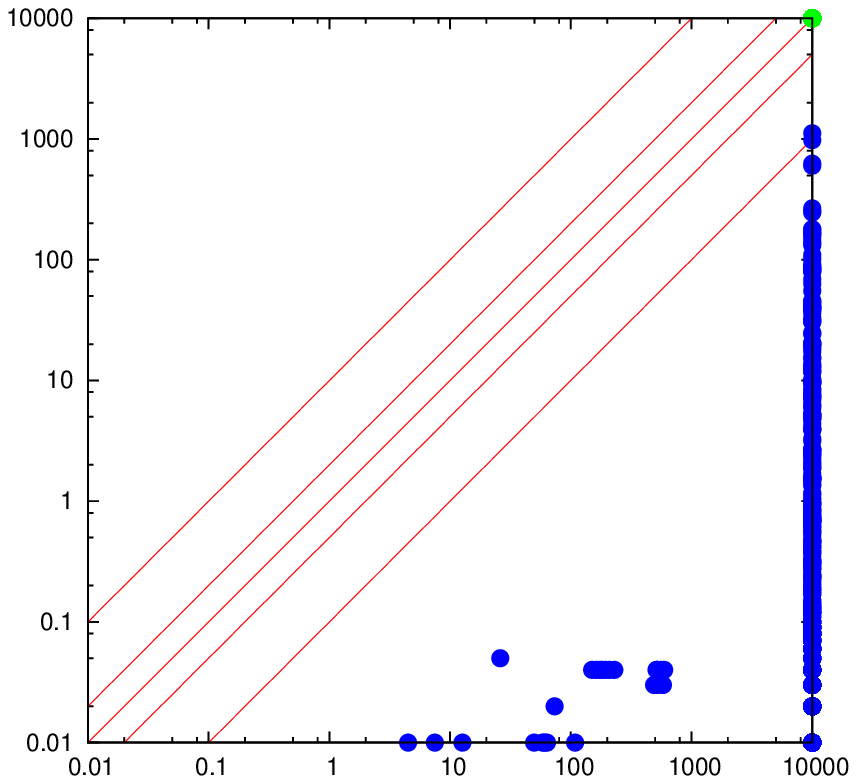} &
    \includegraphics[scale=0.45, bb=126 63 360 285]{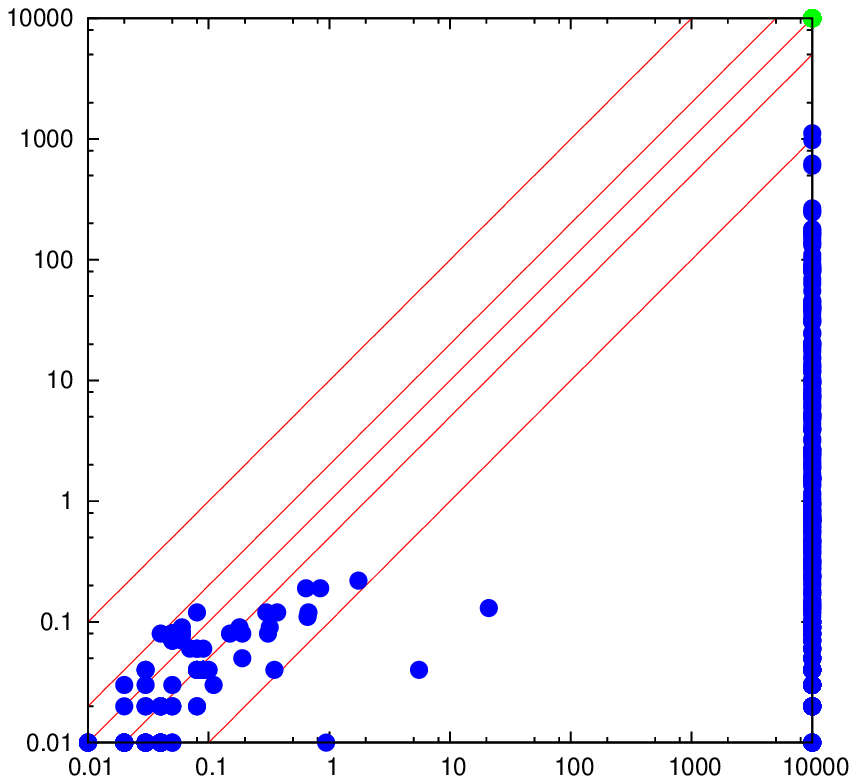} &
    \includegraphics[scale=0.45, bb=126 63 360 285]{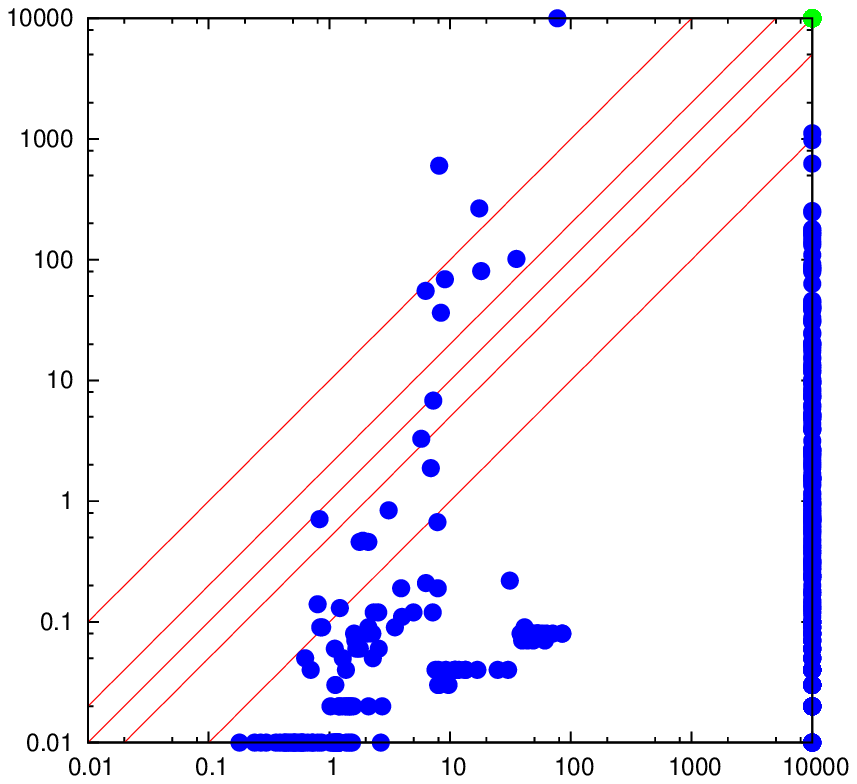} \\[1ex]
    \rotatebox{90}{\hspace{1em}\mathsatceil} &
    \includegraphics[scale=0.45, bb=126 63 360 285]{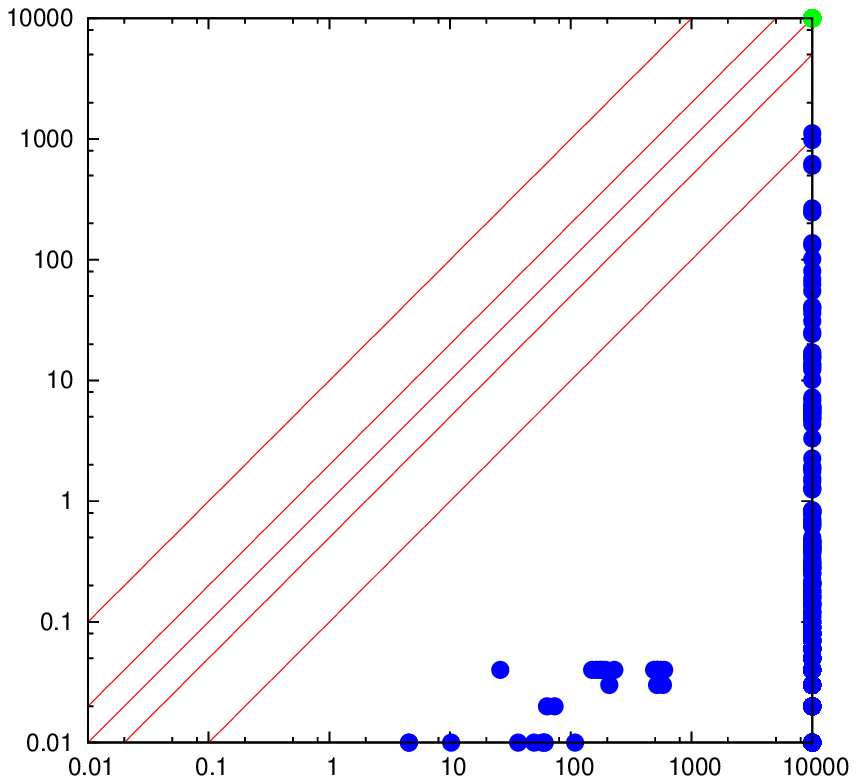} &
    \includegraphics[scale=0.45, bb=126 63 360 285]{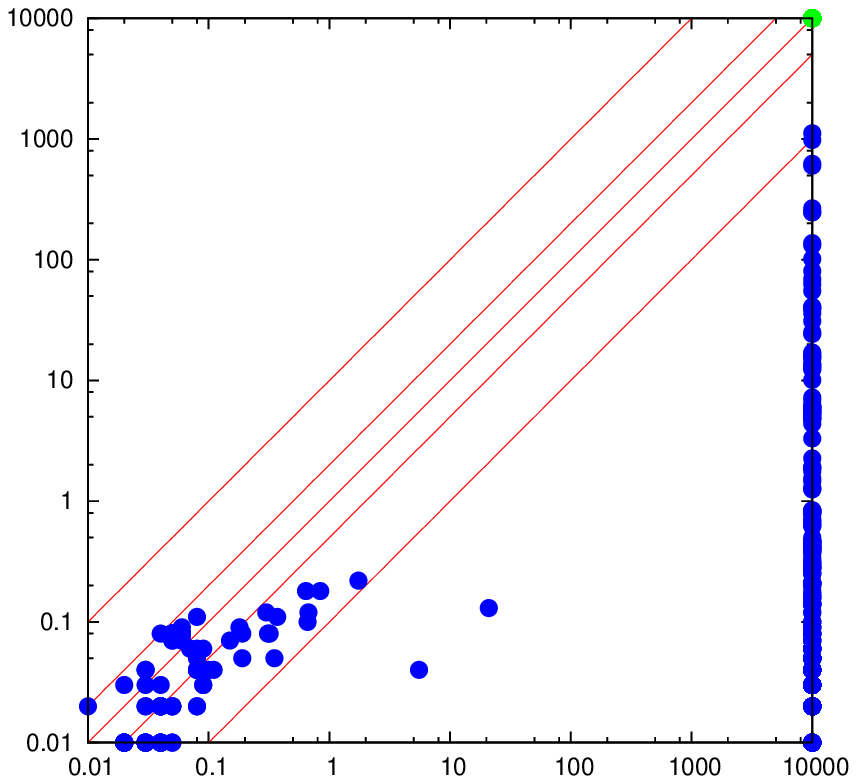} &
    \includegraphics[scale=0.45, bb=126 63 360 285]{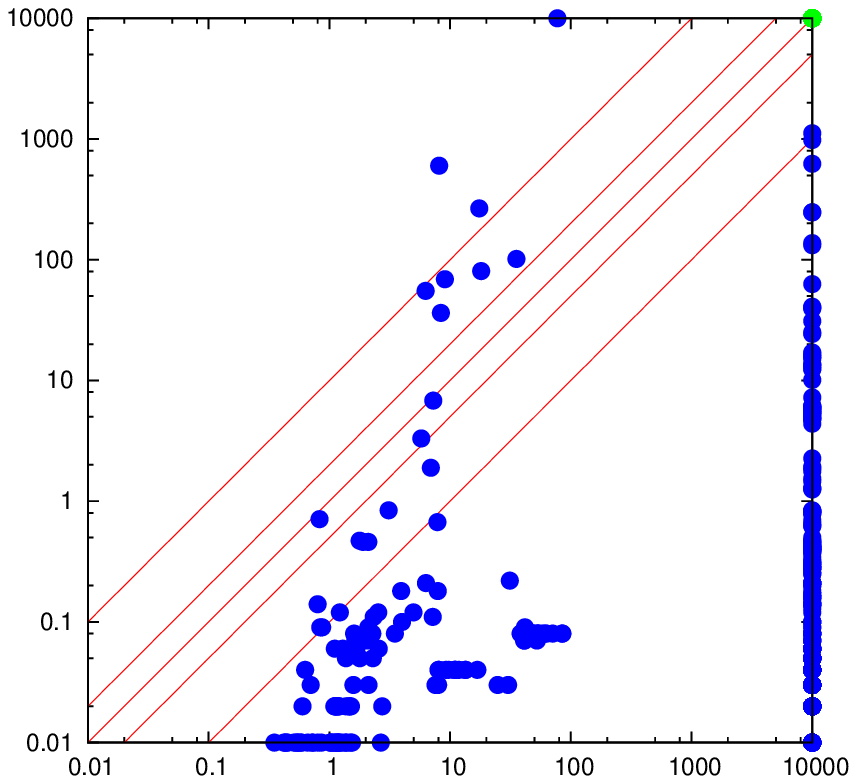}
  \end{tabular}
  \caption{Comparison between \mathsat and the other \laint-interpolating tools, execution time.
  \label{fig:mathsat_vs_others_time}}
\end{figure}

\subsubsection{Comparison with the state-of-the-art tools available}
\label{sec:mathsat_vs_others}

We compare \mathsat
with all the other interpolant generators for \laint 
which are available (to the best of our knowledge):
\iprincess \cite{ijcar10_interpolation},%
\footnote{\url{http://www.philipp.ruemmer.org/iprincess.shtml}}
\opensmtitp \cite{lpar10_interpolation},%
\footnote{\url{http://www.philipp.ruemmer.org/interpolating-opensmt.shtml}}
and \smtinterpol~%
\footnote{\url{http://ultimate.informatik.uni-freiburg.de/smtinterpol/}.
We are not aware of any publication describing the tool.}.
We compare not only the execution times for generating interpolants,
but also the size of the generated formulas 
(measured in terms of number of nodes in their DAG representation).

For \mathsat, we use two configurations:
\mathsatcongr, which produces interpolants with modular equations using the Strengthen rule of \sref{sec:interpolation},
and \mathsatceil, which uses the ceiling function and the Division rule of \sref{sec:itp_ceiling}.

Results on execution times for generating interpolants are reported in Fig.~\ref{fig:mathsat_vs_others_time}.
Both \mathsatcongr and \mathsatceil could successfully generate an interpolant 
for 478 of the 513 interpolation problems (timing out on the others),
whereas \iprincess, \opensmtitp and \smtinterpol 
were able to successfully produce an interpolant in 
62, 192 and 217 cases respectively.
Therefore, \mathsat can solve more than twice as many instances as its closer competitor \smtinterpol,
and in most cases with a significantly shorter execution time (Fig.~\ref{fig:mathsat_vs_others_time}).

\begin{figure}[t]
  \begin{tabular}{cccc}
    & \iprincess & \hspace{1em}\opensmtitp & \hspace{2em}\smtinterpol \\
    \rotatebox{90}{\hspace{0.5em}\mathsatcongr} &
    \includegraphics[scale=0.45, bb=126 63 360 285]{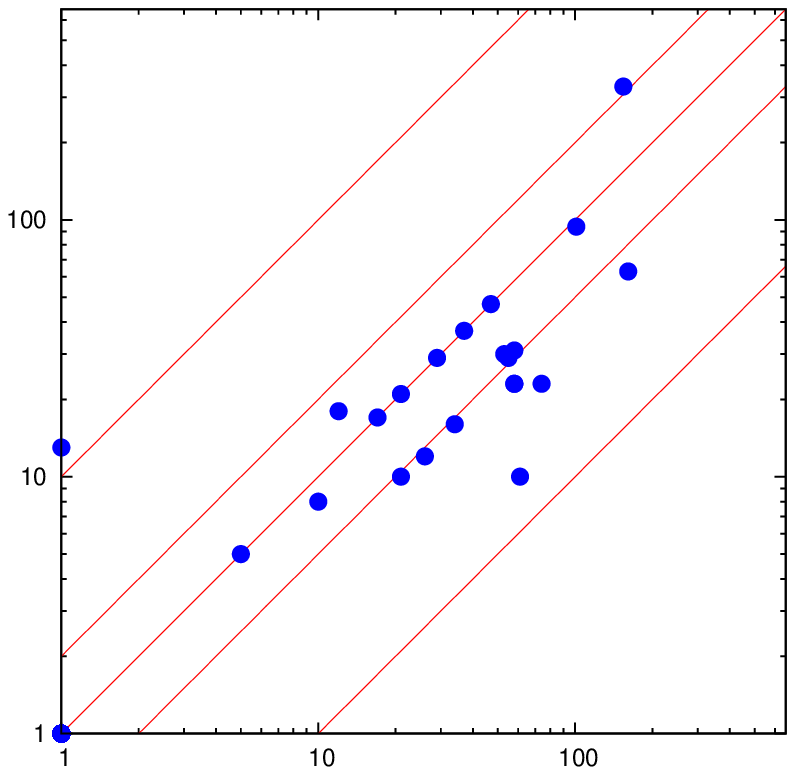} &
    \includegraphics[scale=0.45, bb=126 63 360 285]{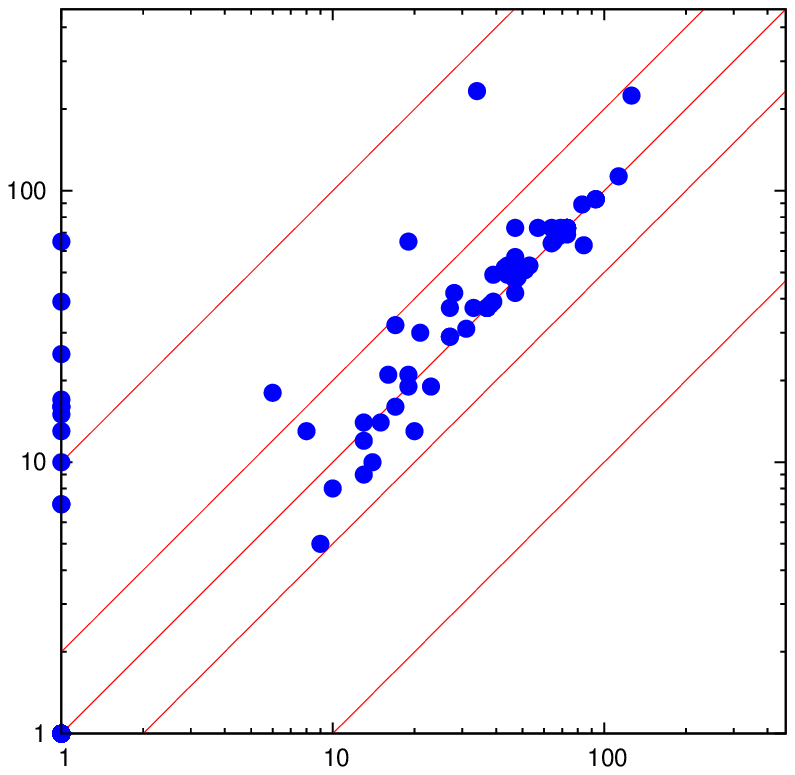} &
    \includegraphics[scale=0.45, bb=126 63 360 285]{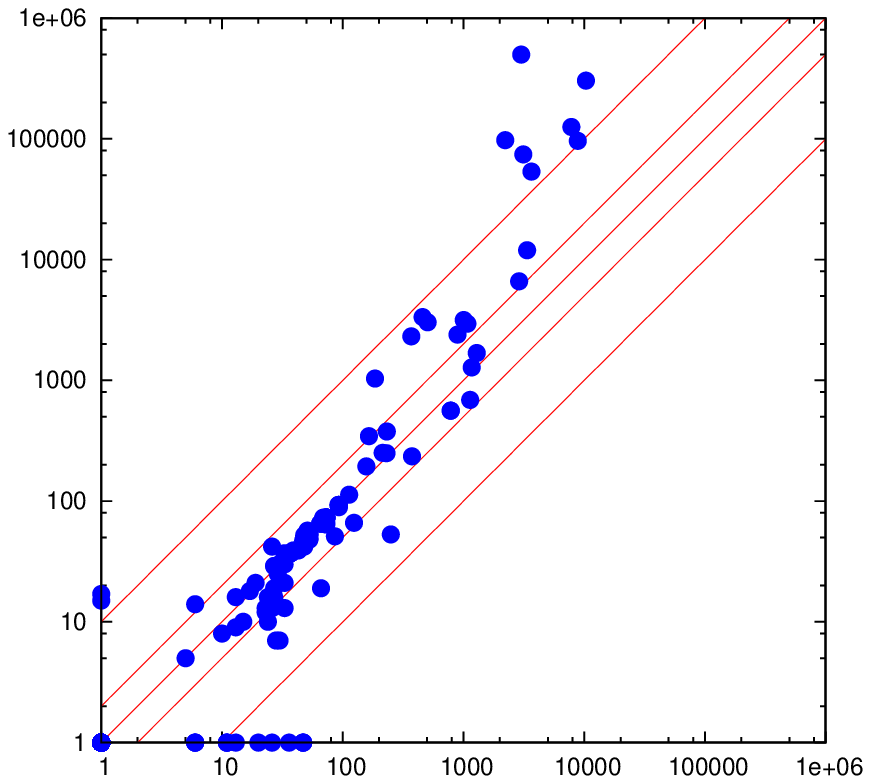} \\[1ex]
    \rotatebox{90}{\hspace{1em}\mathsatceil} &
    \includegraphics[scale=0.45, bb=126 63 360 285]{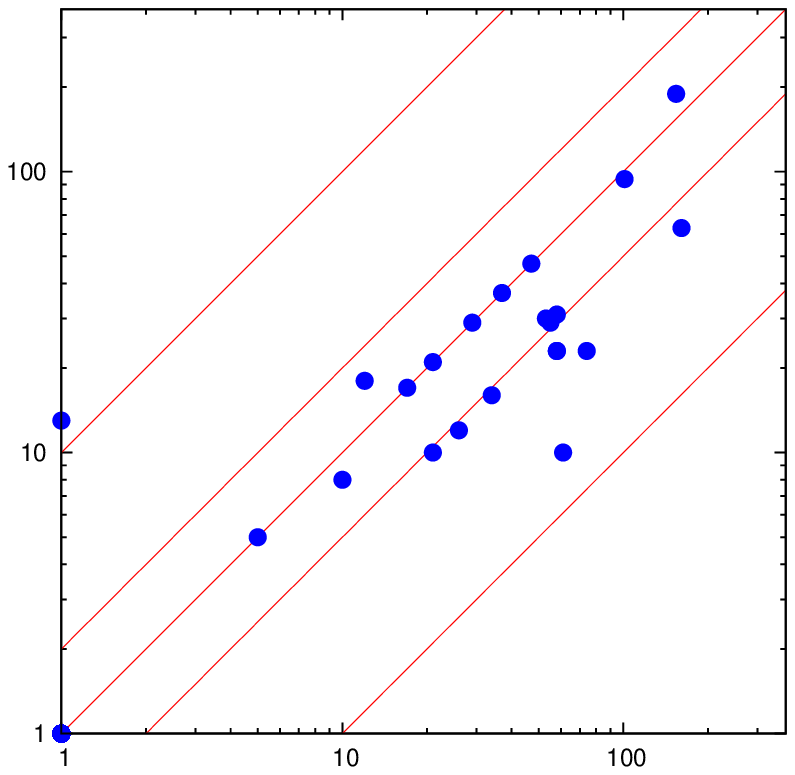} &
    \includegraphics[scale=0.45, bb=126 63 360 285]{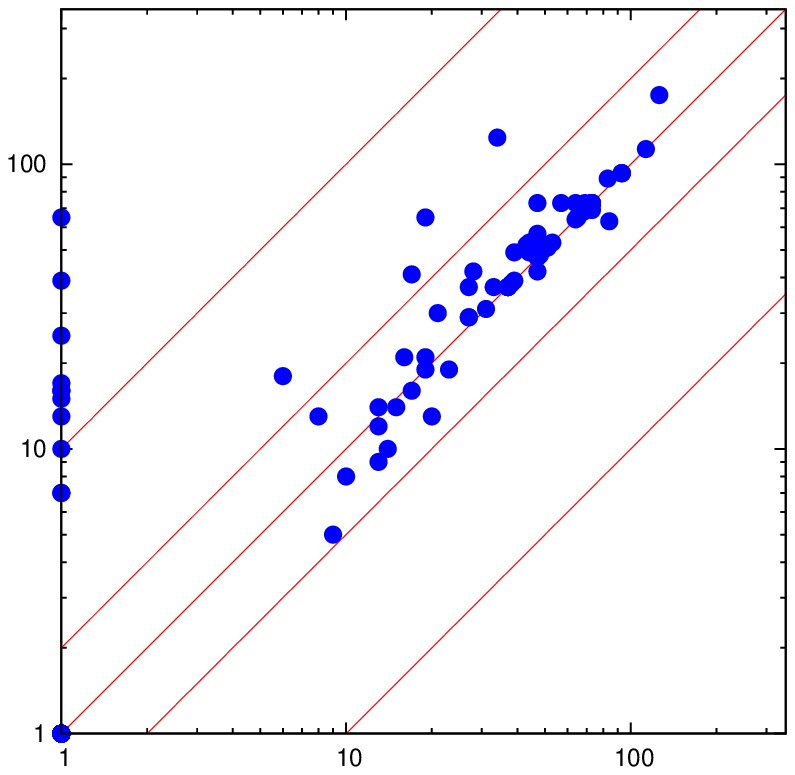} &
    \includegraphics[scale=0.45, bb=126 63 360 285]{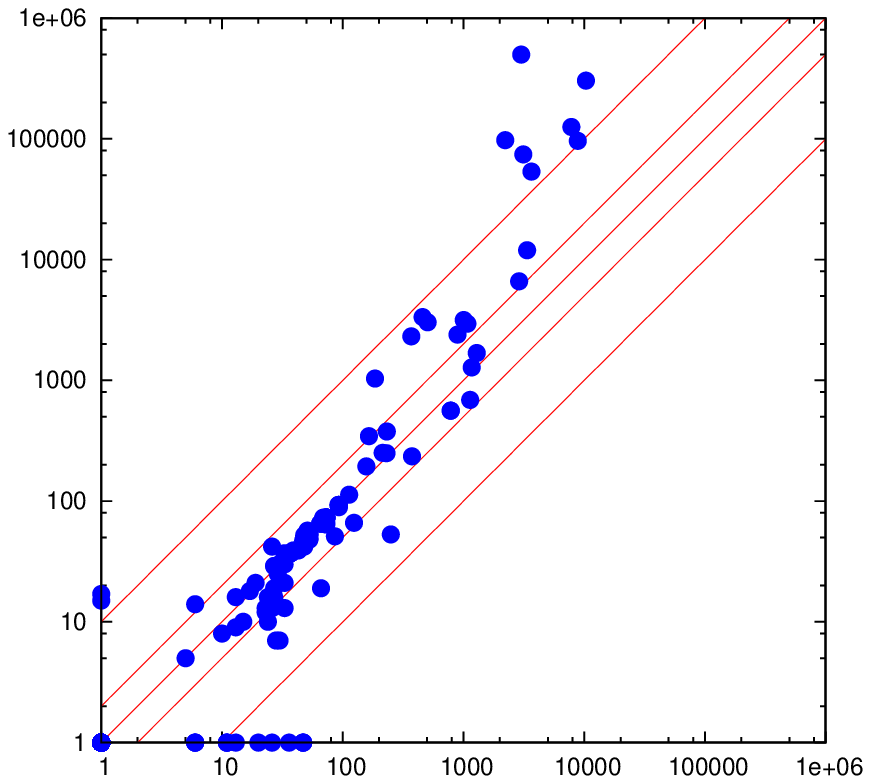}
  \end{tabular}
  \caption{Comparison between \mathsat and the other \laint-interpolating tools, interpolants size 
(measured in number of nodes in the DAG of the interpolant).
  \label{fig:mathsat_vs_others_size}
{(See also footnote~\ref{footnote:outliers}.)}}
\end{figure}

For the subset of instances which could be solved by at least one other tool,
therefore, the two configurations of \mathsat seem to perform equally well.
The situation is the same also when we compare the sizes of the produced interpolants,
measured in number of nodes in a DAG representation of formulas.
Comparisons on interpolant size are reported in Fig.~\ref{fig:mathsat_vs_others_size},
which shows that, on average, the interpolants produced by \mathsat 
are comparable to those produced by other tools.
In fact, there are some cases in which \smtinterpol produces significantly-smaller interpolants,
but we remark that \mathsat can solve 261 more instances than \smtinterpol.%
\footnote{\label{footnote:outliers}
The plots {of Fig.~\ref{fig:mathsat_vs_others_size}} show also some apparently-strange outliers in the comparison with \opensmtitp.
A closer analysis revealed that those are instances for which \opensmtitp was able to detect that the inconsistency of $A \land B$ was due solely to $A$ or to $B$, and thus could produce a trivial interpolant $\bot$ or $\top$,
whereas the proof of unsatisfiability produced by \mathsat involved both $A$ and $B$.
An analogous situation is visible also in the comparison between \mathsat and \smtinterpol, this time in favor of \mathsat.
}

\begin{figure}[t]
  \begin{tabular}{ccc|c}
    & Execution Time & Interpolants Size & Execution Time \\
    \rotatebox{90}{\hspace{1em}\mathsatceil} &
    \includegraphics[scale=0.45, bb=126 63 360 285]{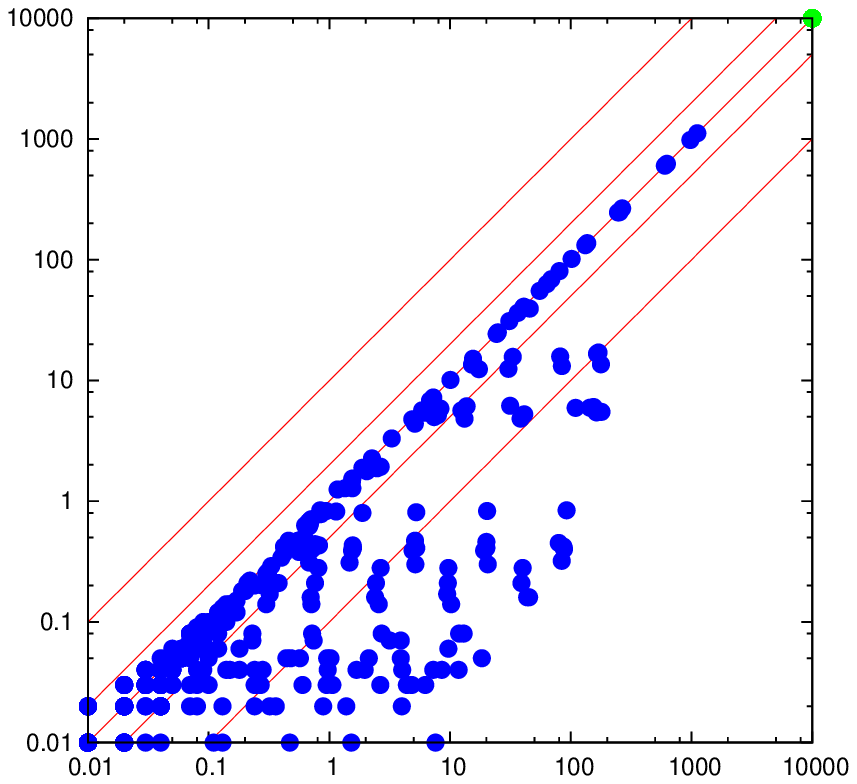} \hspace{0.5ex} &
    \includegraphics[scale=0.45, bb=126 63 360 285]{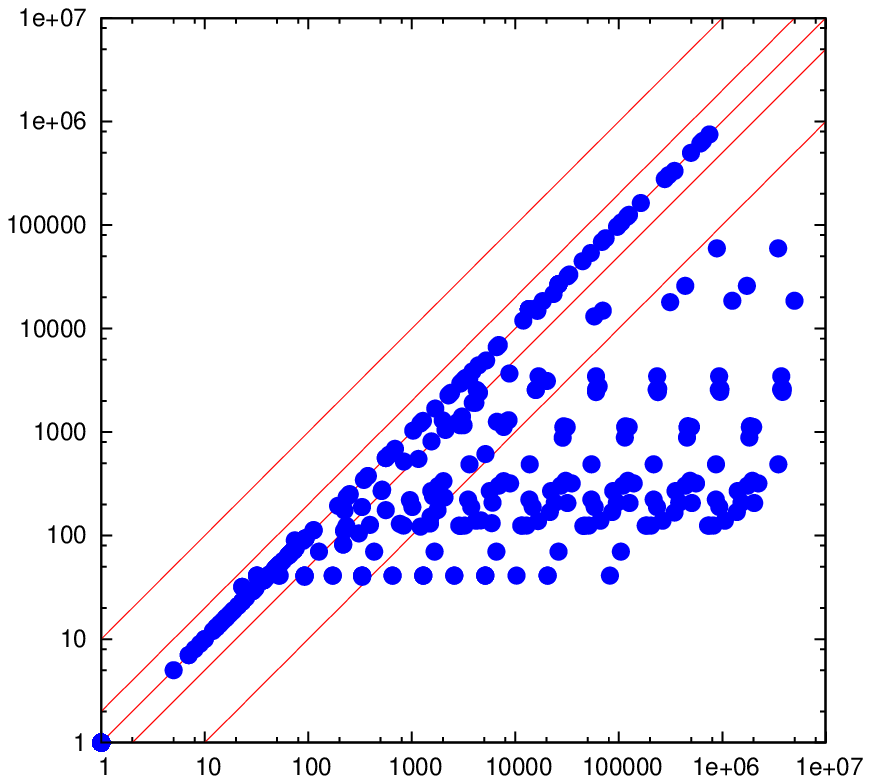} \hspace{0.5em} &
    \includegraphics[scale=0.45, bb=126 63 360 285]{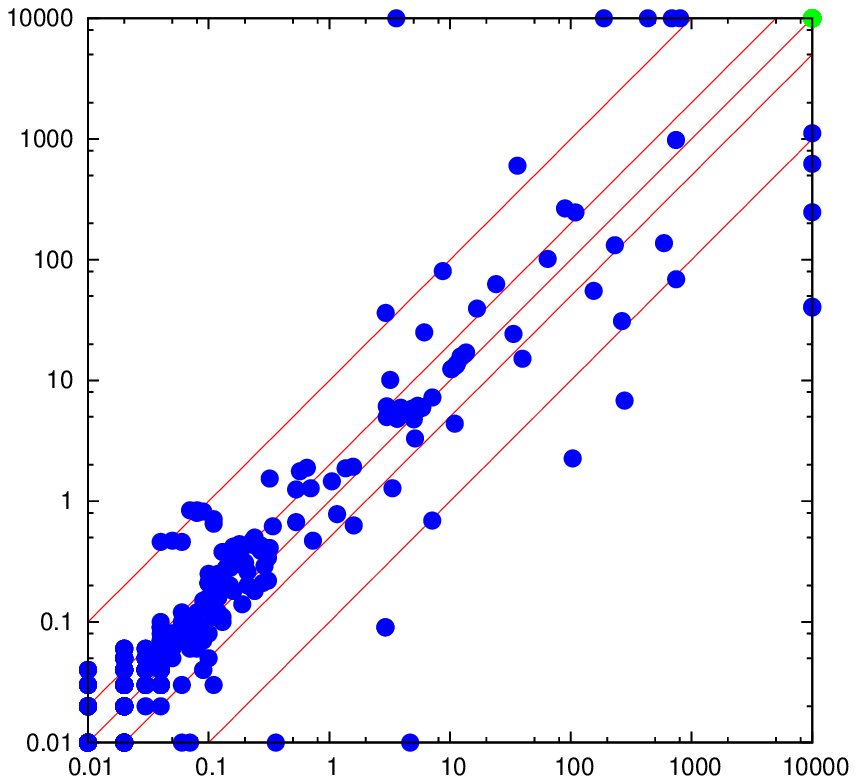} \\
    & \mathsatcongr & \hspace{1em}\mathsatcongr & \hspace{0.5ex}\mathsat w/o interpolation \\
    & \multicolumn{2}{c|}{\textbf{(a)}} & \textbf{(b)}
  \end{tabular}
  \caption{
    (a) Comparison between \mathsatcongr and \mathsatceil configurations for interpolation. 
    (b) Execution time overhead for interpolation with \mathsatceil.
  \label{fig:mathsat_vs_itself}}
\end{figure}

The differences between \mathsatcongr and \mathsatceil become evident 
when we compare the two configurations directly.
The plots in Fig.~\ref{fig:mathsat_vs_itself}(a)
show that \mathsatceil is dramatically superior to \mathsatcongr,
with gaps of up to two orders of magnitude in execution time,
and up to four orders of magnitude in the size of interpolants.
Such differences are solely due to the use of the ceiling function in the generated interpolants,
which prevents the blow-up of the formula wrt. the size of the proof of unsatisfiability.
Since most of the differences between the two configurations occur in benchmarks
that none of the other tools could solve, 
the advantage of using ceilings was not visible in 
Figs.~\ref{fig:mathsat_vs_others_time} and \ref{fig:mathsat_vs_others_size}.

Finally, in Fig.~\ref{fig:mathsat_vs_itself}(b) we compare the execution time of
producing interpolants with \mathsatceil against the solving time of \mathsat
with interpolation turned off.
The plot shows that the restriction on the kind of extended branch-and-bound 
lemmas generated when computing interpolants (see \sref{sec:interpolation_bb})
can have a significant impact on individual benchmarks. 
However, on average \mathsatceil is not worse than the ``regular'' \mathsat,
and the two can solve the same number of instances,
in approximately the same total execution time.

\subsection{Experiments on model checking problems}

\begin{table}[t]
  \begin{scriptsize}
    \centering
  \caption{Experimental results on model checking problems.
    \label{tab:results_itp_mc}}

  \begin{tabular}{l|rrrr}
    \multicolumn{5}{c}{} \\
    & \multicolumn{4}{|c}{{\bf Results (num. of queries / execution time)}} \\
                             & {\bf \mathsatceil} & {\bf \mathsatcongr} & {\bf \mathsat-\textsc{noEQ}} &  {\bf \mathsat-\textsc{noBB}} \\
    \hline
    {\bf byte\_add\_1}       &  9 / 1.05          &    9 / 1.06         & 9 / 1.00                     &  9 / 1.02                     \\
    {\bf byte\_add\_2}       &  13 / 2.33         &    13 / 2.36        & 13 / 2.27                    &  13 / 2.40                    \\
    {\bf byte\_add\_3}       &  52 / 97.68        &    52 / 91.71       & T.O.                         &  52 / 97.83                   \\
    {\bf byte\_add\_4}       &  28 / 27.77        &    28 / 28.38       & T.O.                         &  28 / 28.06                   \\
    {\bf jain\_1}            &  8 / 0.04          &    8 / 0.04         & T.O.                         &  8 / 0.04                     \\
    {\bf jain\_2}            &  8 / 0.06          &    8 / 0.06         & T.O.                         &  8 / 0.05                     \\
    {\bf jain\_4}            &  7 / 0.06          &    7 / 0.06         & T.O.                         &  7 / 0.05                     \\
    {\bf jain\_5}            &  42 / 0.84         &   42 / 0.82         & 42 / 0.88                    &  42 / 0.81                    \\
    {\bf jain\_6}            &  7 / 0.06          &    7 / 0.06         & T.O.                         &  7 / 0.06                     \\
    {\bf jain\_7}            &  8 / 0.08          &    8 / 0.08         & T.O.                         &  8 / 0.07                     \\
    {\bf num\_conversion\_1} &  51 / 13.33        &     51 / 13.02      & 51 / 21.04                   &  51 / 12.83                   \\
    {\bf num\_conversion\_2} &  T.O.              & T.O.                & T.O.                         &  T.O.                         \\
    {\bf num\_conversion\_3} &  43 / 5.40         &  43 / 5.02          & 43 / 5.78                    &  43 / 4.99                    \\
    {\bf num\_conversion\_4} &  52 / 19.03        &  53 / 19.81         & T.O.                         &  52 / 17.45                   \\
    {\bf num\_conversion\_5} &  47 / 8.63         &  47 / 7.87          & T.O.                         &  47 / 7.72                    \\
    \hline
  \end{tabular}

~\\[2em]

  \begin{tabular}{l|rrrr}
    & \multicolumn{4}{|c}{{\bf Results (num. of queries / execution time)}} \\
                             & {\bf \mathsat-\textsc{noEQ-noBB}} & {\bf \smtinterpol} & {\bf \iprincess} & {\bf \opensmtitp} \\    
    \hline
    {\bf byte\_add\_1}       & 9 / 1.60                          & 9 / 46.48          & T.O.             & 1 / 0.073         \\
    {\bf byte\_add\_2}       & 13 / 3.03                         & 9 / 48.37          & T.O.             & 1 / 0.073         \\
    {\bf byte\_add\_3}       & 52 / 111.89                       & ERR                & T.O.             & BAD               \\
    {\bf byte\_add\_4}       & 28 / 44.19                        & ERR                & T.O.             & BAD               \\
    {\bf jain\_1}            & T.O.                              & 7 / 2.15           & 8 / 23.44        & BAD               \\
    {\bf jain\_2}            & T.O.                              & BAD                & 6 / 20.12        & BAD               \\
    {\bf jain\_4}            & T.O.                              & 7 / 2.93           & 9 / 55.05        & BAD               \\
    {\bf jain\_5}            & 42 / 0.81                         & BAD                & T.O.             & BAD               \\
    {\bf jain\_6}            & T.O.                              & BAD                & 7 / 28.96        & BAD               \\
    {\bf jain\_7}            & T.O.                              & BAD                & T.O.             & BAD               \\
    {\bf num\_conversion\_1} & 51 / 11.44                        & BAD                & T.O.             & BAD               \\
    {\bf num\_conversion\_2} & T.O.                              & ERR                & T.O.             & BAD               \\
    {\bf num\_conversion\_3} & 43 / 5.36                         & ERR                & T.O.             & BAD               \\
    {\bf num\_conversion\_4} & 60 / 37.86                        & BAD                & T.O.             & BAD               \\
    {\bf num\_conversion\_5} & 47 / 8.10                         & ERR                & T.O.             & BAD               \\
    \hline
    \multicolumn{5}{p{14cm}}
    {{\bf \textsc{Key:}} T.O.: time-out (300 seconds);
  ERR: internal error/crash of the interpolating solver; BAD: wrong interpolant produced.}
  \end{tabular}
\end{scriptsize}
\end{table}

{
In the second part of our experimental analysis,
we evaluate the performance of \mathsat and all the other interpolant generators for \laint
described above (\S\ref{sec:mathsat_vs_others})
when used in an interpolation-based model checking context.
In particular,
we have implemented the original interpolation-based model checking algorithm of McMillan \cite{interpolation-mc-sat},
and applied it to the verification of some transition systems 
generated from simple sequential C programs, 
using \laint as a background theory.%
\footnote{Both the implementation and the benchmarks are available upon request.}
The benchmarks have been taken from the literature on \laint-related interpolation procedures \cite{jain_cav08,griggio-fmcad11}.
We have then run this implementation using each of the solvers above as interpolation engines,
and compared the results in terms of number of instances solved, 
time spent in computing interpolants, and number of calls to the interpolating solvers.
For \mathsat, besides the two configurations \mathsatcongr and \mathsatceil described in the previous section,
we have also tested additional configurations obtained by disabling some of the layers of the \laint-solver described in \S\ref{sec:background_smtsolving}:
\mathsat-\textsc{noEQ} in which we disabled the equality elimination module,
\mathsat-\textsc{noBB} in which we disabled the internal branch and bound module,
and \mathsat-\textsc{noEQ-noBB} in which we disabled both.
}

{
The results are reported in Table~\ref{tab:results_itp_mc}.
They clearly show that both \mathsatceil and \mathsatcongr outperform the other tools
also when applied in a model checking context.
Moreover, it is interesting to observe the following:
\begin{iteMize}{$\bullet$}
\item for these particular benchmarks, \mathsatcongr and \mathsatceil seem to be substantially equivalent: 
the only significant difference is in the {\tt num\_conversion\_4} benchmark, 
in which \mathsatceil leads to a slightly faster convergence, requiring one interation less;
\item the equality elimination layer seems to be very important, 
and disabling it leads to a dramatic decrease in performance;
\item somewhat surprisingly, 
the decrease in performance due to the disabling of the equality elimination module can be mitigated by disabiling \emph{also} the internal branch and bound module.
We attribute this to the different ``quality'' of the interpolants generated,
which seems to be somehow ``better'' for \mathsat-\textsc{noEq-noBB} than for \mathsat-\textsc{noEQ}.
However, we remark that the notion of ``quality'' of interpolants is still vague and unclear,
and in particular we are not aware of any satisfactory characterization of it in the literature.
Investigating the issue more in depth could be part of interesting future work.
\end{iteMize}
}

\section{Related Work}
\label{sec:related}

The general algorithm for interpolation in \smtt was given by McMillan in \cite{mcmillan_interpolating_prover},
together with algorithms for sets of literals in the theories \euf, \larat and their combination. 
Algorithms for other theories and/or alternative approaches are presented in
\cite{rybalchenko_interp,%
musuvathi_interpolation,%
kroening_interp,%
interpolation_data_structures,%
tocl_interpolation,%
jain_cav08,%
lynch_interpolation,%
tinelli_tacas09,%
tinelli_cade09,%
ijcar10_interpolation,%
lpar10_interpolation%
}.
In particular, \cite{tocl_interpolation,tinelli_tacas09,tinelli_cade09}
explicitly focus on building efficient interpolation procedures on top of 
state-of-the-art \smt solvers.
Efficient interpolation algorithms for the Difference Logic and
Unit-Two-Variables-Per-Inequality fragments of \laint are given in
\cite{tocl_interpolation}.
{Some preliminary work on interpolation on the theory of
  fixed-width bit-vectors is presented in \cite{kroening_interp,griggio-fmcad11}.}
As regards interpolation in the full \laint,
McMillan showed in \cite{mcmillan_interpolating_prover} 
that it is in general not possible to obtain quantifier-free interpolants
(starting from a quantifier-free input)
in the standard signature of \laint 
(consisting of Boolean connectives, integer constants and the symbols
$+, \cdot, \leq, =$).
By extending the signature to contain \emph{modular equalities} 
(or, equivalently, \emph{divisibility predicates})
it is possible to compute quantifier-free \laint interpolants by means of quantifier elimination,
which is however prohibitively expensive in general, both in theory and in practice.
Using modular equalities, 
Jain et al. \cite{jain_cav08} developed polynomial-time interpolation
algorithms 
for linear equations and disequations and for linear modular equations. 
A similar algorithm was also proposed in \cite{lynch_interpolation}.
The work in \cite{ijcar10_interpolation} was the first to present 
an interpolation algorithm for the full \laint 
(augmented with divisibility predicates) not based on quantifier elimination.
Finally, an alternative algorithm, 
exploiting efficient interpolation procedures for \larat and for linear equations in \laint,
has been recently presented in \cite{lpar10_interpolation}.

\section{Conclusions}
\label{sec:concl}

In this article, we have presented a novel interpolation algorithm for \laint
that allows for producing interpolants from arbitrary cutting-plane proofs 
without the need of performing quantifier elimination.
We have also shown how to exploit this algorithm, 
in combination with other existing techniques, 
in order to implement an efficient interpolation procedure on top of a 
state-of-the-art $\smtlaint$-solver, with almost no overhead in search,
and with up to orders of magnitude improvements 
-- both in execution time and in formula size --
wrt. existing techniques for computing interpolants from arbitrary cutting-plane proofs.

\newcommand{\etalchar}[1]{$^{#1}$}

\end{document}

%% file: itp_example_proof.pstex_t
\begin{picture}(0,0)%
\includegraphics{itp_example_proof.pstex}%
\end{picture}%
\setlength{\unitlength}{3947sp}%
\begingroup\makeatletter\ifx\SetFigFont\undefined%
\gdef\SetFigFont#1#2#3#4#5{%
  \reset@font\fontsize{#1}{#2pt}%
  \fontfamily{#3}\fontseries{#4}\fontshape{#5}%
  \selectfont}%
\fi\endgroup%
\begin{picture}(5580,6030)(436,-5851)
\put(451, 14){\makebox(0,0)[lb]{\smash{{\SetFigFont{12}{14.4}{\rmdefault}{\mddefault}{\updefault}{\color[rgb]{0,0,0}\TlemmaOne}%
}}}}
\put(451,-286){\makebox(0,0)[lb]{\smash{{\SetFigFont{12}{14.4}{\rmdefault}{\mddefault}{\updefault}{\color[rgb]{0,0,0}\TlemmaOneb}%
}}}}
\put(3526,-2386){\makebox(0,0)[b]{\smash{{\SetFigFont{12}{14.4}{\rmdefault}{\mddefault}{\updefault}{\color[rgb]{0,0,0}\ResTwo}%
}}}}
\put(1651,-2761){\makebox(0,0)[b]{\smash{{\SetFigFont{12}{14.4}{\rmdefault}{\mddefault}{\updefault}{\color[rgb]{0,0,0}\HypThree}%
}}}}
\put(4201,-3661){\makebox(0,0)[b]{\smash{{\SetFigFont{12}{14.4}{\rmdefault}{\mddefault}{\updefault}{\color[rgb]{0,0,0}\HypFour}%
}}}}
\put(3676,-4261){\makebox(0,0)[b]{\smash{{\SetFigFont{12}{14.4}{\rmdefault}{\mddefault}{\updefault}{\color[rgb]{0,0,0}\ResFour}%
}}}}
\put(4576,-5086){\makebox(0,0)[b]{\smash{{\SetFigFont{12}{14.4}{\rmdefault}{\mddefault}{\updefault}{\color[rgb]{0,0,0}\ResFive}%
}}}}
\put(2926,-5161){\makebox(0,0)[b]{\smash{{\SetFigFont{12}{14.4}{\rmdefault}{\mddefault}{\updefault}{\color[rgb]{0,0,0}\HypSix}%
}}}}
\put(3751,-5836){\makebox(0,0)[b]{\smash{{\SetFigFont{12}{14.4}{\rmdefault}{\mddefault}{\updefault}{\color[rgb]{0,0,0}\Bottom}%
}}}}
\put(2701,-3361){\makebox(0,0)[b]{\smash{{\SetFigFont{12}{14.4}{\rmdefault}{\mddefault}{\updefault}{\color[rgb]{0,0,0}\ResThree}%
}}}}
\put(3076,-661){\makebox(0,0)[b]{\smash{{\SetFigFont{12}{14.4}{\rmdefault}{\mddefault}{\updefault}{\color[rgb]{0,0,0}\HypOne}%
}}}}
\put(2251,-1261){\makebox(0,0)[b]{\smash{{\SetFigFont{12}{14.4}{\rmdefault}{\mddefault}{\updefault}{\color[rgb]{0,0,0}\ResOne}%
}}}}
\put(6001,-4561){\makebox(0,0)[rb]{\smash{{\SetFigFont{12}{14.4}{\rmdefault}{\mddefault}{\updefault}{\color[rgb]{0,0,0}\HypFive}%
}}}}
\put(2251,-1561){\makebox(0,0)[b]{\smash{{\SetFigFont{12}{14.4}{\rmdefault}{\mddefault}{\updefault}{\color[rgb]{0,0,0}\ResOneb}%
}}}}
\put(4051,-1636){\makebox(0,0)[lb]{\smash{{\SetFigFont{12}{14.4}{\rmdefault}{\mddefault}{\updefault}{\color[rgb]{0,0,0}\HypTwo}%
}}}}
\end{picture}%

%% file: itp_example_proof2.pstex_t
\begin{picture}(0,0)%
\includegraphics{itp_example_proof2.pstex}%
\end{picture}%
\setlength{\unitlength}{3947sp}%
\begingroup\makeatletter\ifx\SetFigFont\undefined%
\gdef\SetFigFont#1#2#3#4#5{%
  \reset@font\fontsize{#1}{#2pt}%
  \fontfamily{#3}\fontseries{#4}\fontshape{#5}%
  \selectfont}%
\fi\endgroup%
\begin{picture}(4530,5805)(886,-5851)
\put(3526,-2386){\makebox(0,0)[b]{\smash{{\SetFigFont{12}{14.4}{\rmdefault}{\mddefault}{\updefault}{\color[rgb]{0,0,0}\ResTwo}%
}}}}
\put(4276,-1636){\makebox(0,0)[b]{\smash{{\SetFigFont{12}{14.4}{\rmdefault}{\mddefault}{\updefault}{\color[rgb]{0,0,0}\HypTwo}%
}}}}
\put(4201,-3661){\makebox(0,0)[b]{\smash{{\SetFigFont{12}{14.4}{\rmdefault}{\mddefault}{\updefault}{\color[rgb]{0,0,0}\HypFour}%
}}}}
\put(3676,-4261){\makebox(0,0)[b]{\smash{{\SetFigFont{12}{14.4}{\rmdefault}{\mddefault}{\updefault}{\color[rgb]{0,0,0}\ResFour}%
}}}}
\put(4576,-5086){\makebox(0,0)[b]{\smash{{\SetFigFont{12}{14.4}{\rmdefault}{\mddefault}{\updefault}{\color[rgb]{0,0,0}\ResFive}%
}}}}
\put(2926,-5161){\makebox(0,0)[b]{\smash{{\SetFigFont{12}{14.4}{\rmdefault}{\mddefault}{\updefault}{\color[rgb]{0,0,0}\HypSix}%
}}}}
\put(3751,-5836){\makebox(0,0)[b]{\smash{{\SetFigFont{12}{14.4}{\rmdefault}{\mddefault}{\updefault}{\color[rgb]{0,0,0}\Bottom}%
}}}}
\put(2701,-3361){\makebox(0,0)[b]{\smash{{\SetFigFont{12}{14.4}{\rmdefault}{\mddefault}{\updefault}{\color[rgb]{0,0,0}\ResThree}%
}}}}
\put(3076,-661){\makebox(0,0)[b]{\smash{{\SetFigFont{12}{14.4}{\rmdefault}{\mddefault}{\updefault}{\color[rgb]{0,0,0}\HypOne}%
}}}}
\put(2251,-1261){\makebox(0,0)[b]{\smash{{\SetFigFont{12}{14.4}{\rmdefault}{\mddefault}{\updefault}{\color[rgb]{0,0,0}\ResOne}%
}}}}
\put(901,-211){\makebox(0,0)[lb]{\smash{{\SetFigFont{12}{14.4}{\rmdefault}{\mddefault}{\updefault}{\color[rgb]{0,0,0}\TlemmaOne}%
}}}}
\put(5401,-4561){\makebox(0,0)[rb]{\smash{{\SetFigFont{12}{14.4}{\rmdefault}{\mddefault}{\updefault}{\color[rgb]{0,0,0}\HypFive}%
}}}}
\put(1876,-2761){\makebox(0,0)[b]{\smash{{\SetFigFont{12}{14.4}{\rmdefault}{\mddefault}{\updefault}{\color[rgb]{0,0,0}\HypThree}%
}}}}
\end{picture}%

%% file: laint_architecture.pstex_t
\begin{picture}(0,0)%
\includegraphics{laint_architecture.pstex}%
\end{picture}%
\setlength{\unitlength}{3947sp}%
\begingroup\makeatletter\ifx\SetFigFont\undefined%
\gdef\SetFigFont#1#2#3#4#5{%
  \reset@font\fontsize{#1}{#2pt}%
  \fontfamily{#3}\fontseries{#4}\fontshape{#5}%
  \selectfont}%
\fi\endgroup%
\begin{picture}(6970,4362)(-1067,-3509)
\put(3700,-1896){\makebox(0,0)[b]{\smash{{\SetFigFont{11}{13.2}{\rmdefault}{\mddefault}{\updefault}{\color[rgb]{0,0,0}Internal}%
}}}}
\put(3700,-2052){\makebox(0,0)[b]{\smash{{\SetFigFont{11}{13.2}{\rmdefault}{\mddefault}{\updefault}{\color[rgb]{0,0,0}Branch and Bound}%
}}}}
\put(3775,-2981){\makebox(0,0)[b]{\smash{{\SetFigFont{12}{14.4}{\rmdefault}{\mddefault}{\updefault}{\color[rgb]{0,0,0}Branch and Bound}%
}}}}
\put(3775,-3166){\makebox(0,0)[b]{\smash{{\SetFigFont{12}{14.4}{\rmdefault}{\mddefault}{\updefault}{\color[rgb]{0,0,0}lemmas generator}%
}}}}
\put(5643,584){\makebox(0,0)[rb]{\smash{{\SetFigFont{14}{16.8}{\rmdefault}{\mddefault}{\updefault}{\color[rgb]{0,0,0}\laint-solver}%
}}}}
\put(1662,-190){\makebox(0,0)[b]{\smash{{\SetFigFont{12}{14.4}{\sfdefault}{\mddefault}{\updefault}{\color[rgb]{1,1,1}3}%
}}}}
\put(-571,-1433){\makebox(0,0)[b]{\smash{{\SetFigFont{12}{14.4}{\rmdefault}{\mddefault}{\updefault}{\color[rgb]{0,0,0}\dpll}%
}}}}
\put(5391,-120){\makebox(0,0)[b]{\smash{{\SetFigFont{12}{14.4}{\sfdefault}{\mddefault}{\updefault}{\color[rgb]{1,1,1}2}%
}}}}
\put(1231,-190){\makebox(0,0)[b]{\smash{{\SetFigFont{12}{14.4}{\sfdefault}{\mddefault}{\updefault}{\color[rgb]{1,1,1}1}%
}}}}
\put(2629,-941){\makebox(0,0)[b]{\smash{{\SetFigFont{12}{14.4}{\sfdefault}{\mddefault}{\updefault}{\color[rgb]{1,1,1}2}%
}}}}
\put(1411,-938){\makebox(0,0)[b]{\smash{{\SetFigFont{11}{13.2}{\rmdefault}{\mddefault}{\updefault}{\color[rgb]{0,0,0}\larat-solver}%
}}}}
\put(2609,-1649){\makebox(0,0)[b]{\smash{{\SetFigFont{12}{14.4}{\sfdefault}{\mddefault}{\updefault}{\color[rgb]{1,1,1}3}%
}}}}
\put(2001,-1689){\makebox(0,0)[b]{\smash{{\SetFigFont{11}{13.2}{\rmdefault}{\mddefault}{\updefault}{\color[rgb]{0,0,0}no conflict}%
}}}}
\put(2009,-1953){\makebox(0,0)[b]{\smash{{\SetFigFont{11}{13.2}{\rmdefault}{\mddefault}{\updefault}{\color[rgb]{0,0,0}trail simplifications}%
}}}}
\put(676,-2384){\makebox(0,0)[b]{\smash{{\SetFigFont{12}{14.4}{\sfdefault}{\mddefault}{\updefault}{\color[rgb]{1,1,1}4}%
}}}}
\put(883,-2329){\makebox(0,0)[lb]{\smash{{\SetFigFont{11}{13.2}{\rmdefault}{\mddefault}{\updefault}{\color[rgb]{0,0,0}conflict}%
}}}}
\put(3989,-2504){\makebox(0,0)[b]{\smash{{\SetFigFont{12}{14.4}{\sfdefault}{\mddefault}{\updefault}{\color[rgb]{1,1,1}5}%
}}}}
\put(614,-2954){\makebox(0,0)[b]{\smash{{\SetFigFont{12}{14.4}{\sfdefault}{\mddefault}{\updefault}{\color[rgb]{1,1,1}5}%
}}}}
\put(3628,-2443){\makebox(0,0)[rb]{\smash{{\SetFigFont{11}{13.2}{\rmdefault}{\mddefault}{\updefault}{\color[rgb]{0,0,0}timeout}%
}}}}
\put(853,-2968){\makebox(0,0)[lb]{\smash{{\SetFigFont{11}{13.2}{\rmdefault}{\mddefault}{\updefault}{\color[rgb]{0,0,0}Branch and Bound-lemma}%
}}}}
\put(3895,-507){\makebox(0,0)[b]{\smash{{\SetFigFont{12}{14.4}{\sfdefault}{\mddefault}{\updefault}{\color[rgb]{1,1,1}1}%
}}}}
\put(4887,-761){\makebox(0,0)[b]{\smash{{\SetFigFont{11}{13.2}{\rmdefault}{\mddefault}{\updefault}{\color[rgb]{0,0,0}Diophantine}%
}}}}
\put(4888,-969){\makebox(0,0)[b]{\smash{{\SetFigFont{11}{13.2}{\rmdefault}{\mddefault}{\updefault}{\color[rgb]{0,0,0}equations handler}%
}}}}
\put(4799,-2063){\makebox(0,0)[b]{\smash{{\SetFigFont{12}{14.4}{\sfdefault}{\mddefault}{\updefault}{\color[rgb]{1,1,1}4}%
}}}}
\put(5303,-1679){\makebox(0,0)[b]{\smash{{\SetFigFont{12}{14.4}{\sfdefault}{\mddefault}{\updefault}{\color[rgb]{1,1,1}1}%
}}}}
\put(1471,-439){\makebox(0,0)[lb]{\smash{{\SetFigFont{12}{14.4}{\rmdefault}{\mddefault}{\updefault}{\color[rgb]{0,0,0}conflict}%
}}}}
\put(361,259){\makebox(0,0)[lb]{\smash{{\SetFigFont{11}{13.2}{\rmdefault}{\mddefault}{\updefault}{\color[rgb]{0,0,0}\laint-conflict}%
}}}}
\put(3335,-1003){\makebox(0,0)[b]{\smash{{\SetFigFont{11}{13.2}{\rmdefault}{\mddefault}{\updefault}{\color[rgb]{0,0,0}no conflict}%
}}}}
\put(3352,-1238){\makebox(0,0)[b]{\smash{{\SetFigFont{11}{13.2}{\rmdefault}{\mddefault}{\updefault}{\color[rgb]{0,0,0}equality elimination}%
}}}}
\put(3311,-576){\makebox(0,0)[b]{\smash{{\SetFigFont{11}{13.2}{\rmdefault}{\mddefault}{\updefault}{\color[rgb]{0,0,0}no conflict}%
}}}}
\put(4948,-2369){\makebox(0,0)[b]{\smash{{\SetFigFont{11}{13.2}{\rmdefault}{\mddefault}{\updefault}{\color[rgb]{0,0,0}\laint model}%
}}}}
\put(5172,-325){\makebox(0,0)[rb]{\smash{{\SetFigFont{11}{13.2}{\rmdefault}{\mddefault}{\updefault}{\color[rgb]{0,0,0}conflict}%
}}}}
\put(5114,-1566){\makebox(0,0)[rb]{\smash{{\SetFigFont{11}{13.2}{\rmdefault}{\mddefault}{\updefault}{\color[rgb]{0,0,0}\laint model}%
}}}}
\put(5506,-3068){\makebox(0,0)[b]{\smash{{\SetFigFont{12}{14.4}{\rmdefault}{\mddefault}{\updefault}{\color[rgb]{0,0,0}\satres}%
}}}}
\end{picture}%